# Enhanced charge density wave coherence in a light-quenched, high-temperature superconductor

**Authors:** S. Wandel[1], F. Boschini[2,3,4], E. H. da Silva Neto[5,6,7], L. Shen[1,8], M. X. Na[2,3], S. Zohar[1], Y. Wang[1], S. B. Welch[1], M. H. Seaberg[1], J. D. Koralek[1], G. L. Dakovski[1], W. Hettel[1], M-F. Lin[1], S. P. Moeller[1], W. F. Schlotter[1], A. H. Reid[1], M. P. Minitti[1], T. Boyle[5,6,7], F. He[9], R. Sutarto[9], R. Liang[2,3], D. Bonn[2,3], W. Hardy[2,3], R. A. Kaindl[10,11], D. G. Hawthorn[12], J.-S. Lee[13], A. F. Kemper[14], A. Damascelli[2,3], C. Giannetti[15], J. J. Turner[1,8], G. Coslovich[1]*

**Affiliations:**

[1]Linac Coherent Light Source, SLAC National Accelerator Laboratory, Menlo Park, California 94720, USA

[2]Department of Physics and Astronomy, University of British Columbia, Vancouver, V6T 1Z1, Canada

[3]Quantum Matter Institute, University of British Columbia, Vancouver, V6T 1Z4, Canada

[4]Centre Énergie Matériaux Télécommunications, Institut National de la Recherche Scientifique, Varennes, QC J3X 1S2, Canada

[5]Department of Physics, Yale University, New Haven, Connecticut 06520, USA

[6]Energy Sciences Institute, Yale University, New Haven, Connecticut 06516, USA

[7]Department of Physics, University of California, Davis, California, 95616, USA

[8]Stanford Institute for Materials and Energy Sciences, SLAC National Accelerator Laboratory and Stanford University, Menlo Park, California 94025, USA

[9]Canadian Light Source, Saskatoon, Saskatchewan, S7N 2V3, Canada

[10]Department of Physics, Arizona State University, Tempe, AZ 85287, USA

[11]Materials Sciences Division, Lawrence Berkeley National Laboratory, Berkeley, CA 94720, USA

[12]Department of Physics and Astronomy, University of Waterloo, Waterloo, N2L 3G1, Canada

[13]Stanford Synchrotron Radiation Lightsource, SLAC National Accelerator Laboratory, Menlo Park, California 94720, USA

[14]Department of Physics, North Carolina State University, Raleigh, NC 27695, U.S.A.

[15]Department of Mathematics and Physics, Università Cattolica del Sacro Cuore, Brescia, BS I-25121, Italy

*Correspondence to: gcoslovich@slac.stanford.edu.

**Abstract:**

**Superconductivity and charge density waves (CDW) are competitive, yet coexisting orders in cuprate superconductors. To understand their microscopic interdependence, a probe capable of discerning their interaction on its natural length and time scales is necessary. We use ultrafast resonant soft x-ray scattering to track the transient evolution of CDW correlations in $YBa_2Cu_3O_{6+x}$ following the quench of superconductivity by an infrared laser pulse. We observe a non-thermal response of the CDW order characterized by a near doubling of the correlation length within ≈ 1 picosecond of the superconducting quench. Our results are consistent with a model in which the interaction between superconductivity and CDW manifests inhomogeneously through disruption of spatial coherence, with superconductivity playing the dominant role in stabilizing CDW topological defects, such as discommensurations.**

**Main Text:** Unconventional superconductivity often emerges in proximity to other low-energy broken-symmetry states, including antiferromagnetism, spin order, and charge density waves (CDW). These orders exist on similar energy scales and often become "intertwined" (*1*, *2*), manifesting a complex interplay arising from strong correlations. The presence of CDW order – a periodic modulation of the electronic density – is ubiquitous in cuprates (*3-14*), and has been detected by x-ray scattering (*7-11*, *13-15*), scanning tunneling spectroscopy (STS) (*4-6*, *10*, *12*, *16*), and nuclear magnetic resonance (NMR) (*17*, *18*). The periodicity of such CDW is generally incommensurate, however, STS and NMR experiments have revealed locally commensurate CDW patterns (*4*, *5*, *12*, *17*, *18*). CDW order can coexist with superconductivity below the superconducting critical temperature ($T_C$), and its temperature dependence suggests a tendency to compete with superconductivity (*8*, *10*). In $YBa_2Cu_3O_{6+x}$ (YBCO), high-magnetic-fields suppress superconductivity and lead to the enhancement of CDW ordering (*15*, *17*, *19*). More recently, CDW enhancement has been achieved by means of applying uniaxial pressure (*20*), yet dynamical information about the interaction between order parameters has not been accessed so far.

An emerging approach to study intertwined orders is to measure their response to ultrafast photo-excitation in the time domain. Ultrafast laser pulses enable selective quenching of orders, thereby inducing an enhancement of the competing orders (*21-24*). The use of free electron lasers (FEL) enables ultrafast x-ray scattering experiments in charge-ordered systems. However, such studies have thus far centered on the photo-induced melting dynamics of charge-order (*25-29*). In this work, we focus instead on unveiling the interaction between superconductivity and CDW in the time domain. We do so by performing time-resolved resonant soft x-ray scattering on YBCO, tracking the CDW response to a laser-driven ultrafast quench of superconductivity.

To probe electronic modulations connected to the CDW we used an incident photon energy of 931.5 eV, resonant with the Cu $L_3$-edge, and scanned the momentum transfer around wavevector $\boldsymbol{Q}_{\mathrm{CDW}}$ = (0.31,0,1.45) (Fig. 1A). The sample was photo-excited by 100-fs infrared laser pulses, variably delayed in time with respect to the 100-fs x-ray pulses. The doping of the sample (x = 0.67) was selected to be at the apex of the CDW dome (Fig. 1B), maximizing the competition with superconductivity. The experiment was performed in the superconducting phase (< 20 K), and for comparison at $T_C$ = 65 K, where the CDW scattering intensity is maximal [Fig. 1A and Ref. (*8*)]. It has been shown that 800-nm laser pulses strongly perturb superconductivity in cuprates, inducing a non-thermal phase transition within a few hundreds of femtoseconds (*23*, *30*-*33*). Here we kept the fluence of the pump pulses near the minimum needed to quench the superconducting phase (*32*), 0.05 mJ/cm$^2$, such that parasitic local heating is minimized.

The photo-induced response of the CDW state in the time-domain is tracked by probing the peak intensity integrated around $\boldsymbol{Q}_{\mathrm{CDW}}$ at varying time delays with respect to the pump pulse. At $T_C$, the pump induces a prompt resolution-limited decrease of the CDW peak intensity (Fig. 1C), followed by a ≈ 3 ps recovery. This dynamic behavior is similar to the well-known photo-induced melting in other charge ordered systems (*25*-*28*, *34*, *35*). However, below $T_C$ (at 20 K) the pump yields the opposite effect, inducing an enhancement of the CDW peak signal with dramatically different dynamics (Fig. 1E). Whereas at 65 K a relatively high pump fluence (0.16 mJ/cm$^2$) is necessary to observe the prompt CDW melting with good signal-to-noise ratio, at 20 K a low pump fluence (0.05 mJ/cm$^2$) produces a large enhancement of the CDW signal, with a rise time of 1.5 ps, and a considerably longer decay time (7 ps). This peculiar response, appearing only below $T_C$, is clearly linked to the presence of the superconducting condensate. We reiterate that the low fluence pump allows to transiently quench superconductivity, without producing significant thermal effects. The maximum rise of lattice temperature is estimated to be about 20 K and the dashed line in Fig.1E displays the estimated heat-induced changes in reflectivity (*36*). In addition, the CDW enhancement signal saturates at this fluence (Fig. 1D), and for higher fluences (> 0.1 mJ/cm$^2$) the CDW response switches to a fast melting and a slow enhancement, with a fluence-dependent cross-over time (Fig. S3) (*36*). Notably, this enables full optical control over the CDW phase, switching between enhancement and suppression by tuning the laser power.

To better understand these results and draw a direct comparison with the dynamics of the superconducting order, we performed complementary transient optical reflectivity measurements at 800 nm. At this wavelength we expect to observe the well-known spectral weight transfer from the low-energy superconducting gap to interband transitions (*23*). The transient optical reflectivity signal is dominated by the superconducting response, as confirmed by its temperature and fluence dependence (Fig.S4) and by comparison to transient MIR spectroscopy resonant to the superconducting gap on an analogous sample (Fig. S5) (*36*, *37*). The optical reflectivity dynamics show typical features resulting from the quench of the superconducting state – the signal saturates at a critical fluence, $\Phi_C$ (*23*, *30*-*32*) (Fig. 1D) above which a flat-top dynamics develops (Fig. S4) (*33*, *36*). When we normalize the transient reflectivity by its saturated value, we note that at the fluence of 0.05 mJ/cm$^2$ superconductivity is almost completely quenched (> 90%) in the probing volume. Superconductivity recovers on a similar timescale as the CDW (Fig. 1E); however, their early dynamics are clearly distinct and cannot be described by a thermodynamic variable, such as an effective temperature. After subtracting a small residual melting signal, we note that the CDW enhancement happens on a timescale of $\approx$ 1 ps, reaching a maximum around $\approx$ 2 ps (*36*), whereas the quench of the superconductivity is complete within 300 fs. Thus, within the first $\approx$ 2 ps after photo-excitation, we observe the non-thermal dynamical reaction of CDW to the quench of superconductivity.

To obtain a snapshot of the CDW spatial correlations following photo-excitation, we scanned the $H$-direction, $\boldsymbol{Q}$ = ($H$, 0, 1.45), in reciprocal space (Fig. 2) at the delay times where photo-induced variations are maximal (~ 0.5 ps at $T_C$ and ~ 2 ps below $T_C$). At 65 K the CDW peak signal suppression is given by a decrease of integrated intensity (16%), as well as a decrease of correlation length (25%). This response is comparable for example to the ultrafast CDW melting in tri-tellurides (*38*), with the variation in correlation length commonly attributed to the formation of topological defects (*25*, *38*). When the system is cooled to well below $T_C$ (12 K, Fig. 2B), the $H$-scan reveals a considerable CDW peak enhancement of $\approx$ 120% following photo-excitation. The CDW peak intensity increase is mostly caused by to the dramatic peak narrowing, indicating a $\approx$ 90% increase of the correlation length, from (36 ± 3) Å to (69 ± 5) Å, i.e., almost a doubling in size of coherent CDW domains. Concurrently, the integrated intensity increases, but by a smaller amount (10-20 %). We also note a ≈ 0.003 r.l.u. momentum shift at 12 K. This shift is not present in the melting case (65 K), suggesting that it is not related to heating or to CDW sliding effects directly induced by light (*35*).

The nature of the non-thermal CDW state becomes evident when comparing the correlation lengths along the a-axis ($\xi_a$) and the integrated intensities ($I_{\mathrm{int}}$) obtained after photo-excitation with a conventional equilibrium synchrotron-based temperature-dependence study (Fig. 2, C-D). The CDW intensities and $\xi_a$ at 65 K and 12 K obtained at the FEL before photo-excitation (blue symbols) are in line with the equilibrium characterization (black symbols). On the other hand, the dynamical CDW data (red symbols) clearly do not follow the equilibrium temperature dependence, i.e., cannot be described by an effective temperature scenario. The increase in correlation length is well beyond what can be achieved at equilibrium, whereas the integrated intensity increase remains small compared to the temperature dependent swing. These data clearly corroborate that the non-thermal state created by the 800-nm laser pulse has unique properties.

The ability to measure both CDW and superconductivity on the same sample with similar excitation conditions provides a unique perspective on the dynamics of the intertwined orders. In the following, we discuss three possible scenarios for the superconductivity-CDW interaction. First, we consider the case of homogeneous and coexisting orders, which we interpret within the framework of time-dependent Ginzburg-Landau (TDGL) theory, as discussed in other systems (*22*, *39*). TDGL predicts that for homogenous and competitive orders, the CDW order parameter amplitude would increase on picosecond timescale, driven by the quench of superconductivity (*36*). This scenario is depicted in Fig. 3A and the corresponding CDW scattering peak shows a dramatic increase of integrated intensity [cfr. (*24*, *36*)]. This is in stark disagreement with our data, where the signal is dominated by a change of correlation length instead. We thus conclude that a simple competition model, assuming homogeneous and coexisting orders, is incompatible with our results.

A second scenario is that of phase-separated strongly competitive orders. In this case we consider well separated CDW and superconducting domains [see e.g., Ref. (*11*)], with an average spacing between neighboring CDW domains larger than the CDW periodicity (> 1 nm). In this scenario, depicted in Fig. 3B, the correlation length $\xi_a$ coincides with the size of the CDW domains, which expand in response to the suppression of neighboring superconducting domains. This inflation leads to an increase of integrated intensity $I_{\text{int}}$ thanks to the expanding CDW filling factor, where $I_{\text{int}} \propto {\xi_a}^2$. Such scaling is incompatible with our data, ruling out domain expansion as the root cause of the dramatic correlation length increase we observe within ≈ 2 ps after photo-excitation. Even in the case of a unidirectional expansion of domains, where the filling factor scales linearly with $\xi_a$, hence $I_{\text{int}} \propto \xi_a$ , this scenario remains incompatible with the data (see Fig. 3B and (*36*)).

To explain our observations, we describe a third hypothesis, illustrated in Fig. 3C. Before photo-excitation, the spatial coherence of the CDW domain is disrupted by a superconductivity-induced defect, stabilized by an interstitial superconducting region within the domain. Similar situations have been discussed theoretically (*40*-*42*), in the context of CDW dislocations primarily caused by quenched disorder, with superconductivity forming opportunistically around them. However, in our experiment, the light-quench of the superconducting condensate directly causes a reaction of the CDW within ≈ 1 ps. Extrinsic CDW defects, i.e., pinned to disorder, typically relax on much longer timescales, of hundreds of picoseconds (*43*) or longer (*44*), with light-driven dynamics of oxygen ordering taking place on timescales of the order of minutes (*45*). Furthermore, the depinning process is thermally activated (*42*, *44*, *46*), which is clearly incompatible with our observation of a faster defect removal dynamics at 20 K (≈ 1 ps) as compared to 65 K (> 3 ps). We thus conclude that our data are consistent with superconductivity playing the leading role in shaping the inhomogeneous CDW landscape by creating intrinsic defects, which rapidly disappear as superconductivity is quenched. To further corroborate this point, we remark that vacancies caused by quenched disorder primarily affect the CDW integrated intensity $I_{\text{int}}$, as reported in Ref.(*47*) and confirmed by our simulations (*36*).

For comparison, we note that high-magnetic-field and uniaxial pressure experiments can explore the effects of quasi-static suppression of superconductivity. These works report an increase of the integrated intensity in addition to the increase of correlation length (*15*, *19*, *20*). We interpret this difference in terms of the slower timescale of those external stimuli (> $10^{-6}$ s), which in turn affect both extrinsic and intrinsic defects. Our experiment instead, by using ultrafast light-excitation, can isolate the dynamics of the rapidly reacting intrinsic defects, which are primarily caused by superconductivity and not pinned to disorder.

In terms of the microscopic character of these superconductivity-induced defects, we reiterate that non-topological point-like defects in the CDW pattern, such as vacancies, can be excluded, as their photo-induced removal should primarily affect the integrated intensity $I_{\text{int}}$ (*36*), in disagreement with the data. We thus consider the case of topological defects, such as CDW dislocations and discommensuration lines (sketched in Fig.3C), whose presence is well known in cuprates (*12*, *16*, *18*) and can affect CDW coherence beyond small local perturbations. We find that the removal of discommensuration lines following the photo-quench of superconductivity captures the salient aspects of the data (Fig. 3C), where the correlation length $\xi_a$ nearly doubles, without a significant increase in integrated intensity (*36*). The sudden photo-quench of superconductivity leads to the annihilation of this topological defect, thus reestablishing phase coherence. Within this model, the ≈ 0.003 r.l.u. momentum shift reported in Fig. 2B can be explained in terms of a ≈ 1% increase in periodicity following defect annihilation and subsequent expansion of the CDW pattern within the domain.

The presence of periodic discommensurations lines could reconcile the experimental observation of an incommensurate CDW wavevector (*6-11*, *13-16*), with reports of local commensurate order, detected in Bi2212 (*4*, *5*, *12*) and YBCO (*17*, *18*). Importantly, our data are consistent with an active role of superconductivity in the formation and stabilization of discommensurations, suggestive of a strong coupling between such defects and superconductivity, as recently discussed in transition-metal dichalcogenides (*48*). Within this scenario, we expect superconductivity to be enhanced along the discommensuration lines, as depicted in Fig.3C, forming a corresponding spatially-modulated state, in line with the observation of pair-density-waves in Bi-based superconductors (*49*, *50*). While we cannot determine the periodicity of this defect at the moment, we estimate an average spacing of at least ≈ 3-4 nm between discommensurations (*36*).

Importantly, in our experiment superconductivity appears to disrupt the spatial coherence within a CDW domain. This differs markedly from a phase separated state for strongly competitive orders and corroborates the intertwining between superconductivity and CDW. In this context, we also note that an ultrafast sliding of CDW is required to restore a common phase pattern across the CDW domain within ≈ 1 ps. The timescale of this CDW condensate motion is compatible with the relaxation dynamics of periodic lattice distortions (*34*), with an estimated speed of $\approx 3\text{x}10^5$ cm/s comparable to the speed of sound in YBCO (*51*).

In summary, by utilizing an ultrafast probe of the CDW order parameter following the optical quench of superconductivity, we observe their dynamical interaction. The results suggest that superconductivity disrupts the CDW spatial coherence within CDW domains, by stabilizing topological defects, such as discommensurations. While further studies are needed to unequivocally identify the microscopic nature of these defects, this approach establishes opportunities to study competing orders complementary to current steady-state methods, such as high-fields and uniaxial pressure experiments. By observing the early state following the quench of one of the order parameters, our strategy allows sensitive detection of spatial patterns intrinsically related to the interaction between intertwined orders.

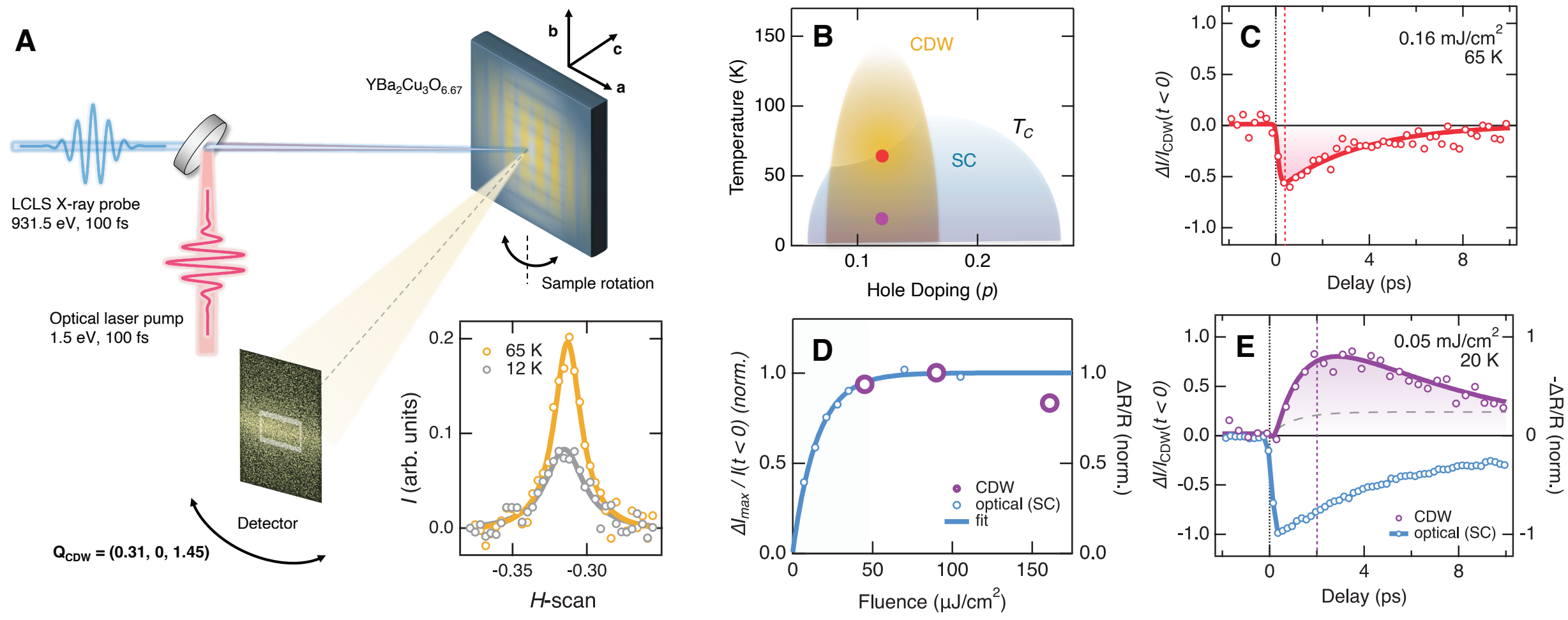

**Fig. 1. Experimental conditions and time-domain results. (A)** Diagram of the experiment. Optical laser pulses perturb the sample and x-rays pulses probe the CDW peak. The momentum space is explored by rocking the sample. Both a 2D area detector and a point detector (gray rectangle) were used to collect the scattered light. Two representative *H*-scans, $\boldsymbol{Q}$ = (*H*, 0, 1.45), are shown in the inset, after subtraction of fluorescence background, at a temperature of 12 and 65 K (*36*). **(B)** Schematic phase diagram of $YBa_2Cu_3O_{6+x}$ showing the two temperatures of interest, within (20 K) and just above (65 K) the superconducting dome for x=0.67 (hole doping *p*=0.12). **(C)** CDW signal at 65 K as a function of delay time between pump and probe pulses, with a laser pump fluence of 0.16 mJ/cm$^2$. The solid line is a fit to exponential relaxation dynamics. **(D)** Fluence dependence of the normalized maximum transient optical signal (blue circles) and x-ray scattering photo-induced enhancement (purple circles) at 20 K (*36*). The blue line is a fit with an exponential function representing saturation of the superconducting response. The gray area highlights fluences below the saturation point. **(E)** Optical (blue circles) and x-ray scattering (purple circles) signals measuring the superconducting and CDW dynamics, respectively, at 20 K with a laser pumping fluence of 0.05 mJ/cm$^2$. The solid purple line indicates a fit of the x-ray scattering dynamics, where the largest component is a positive enhancement of the CDW signal (*36*). The gray dashed line represents the estimate of quasi-thermal heating effects (*36*). To match the fluence value, the optical data are obtained by averaging the two closest curves from the full fluence dependence characterization (Fig.S4). In **(C)** and **(E)** the vertical dotted lines mark time-zero, and the vertical dashed lines mark the delays considered in Fig.2, at which photo-induced variations are maximal.

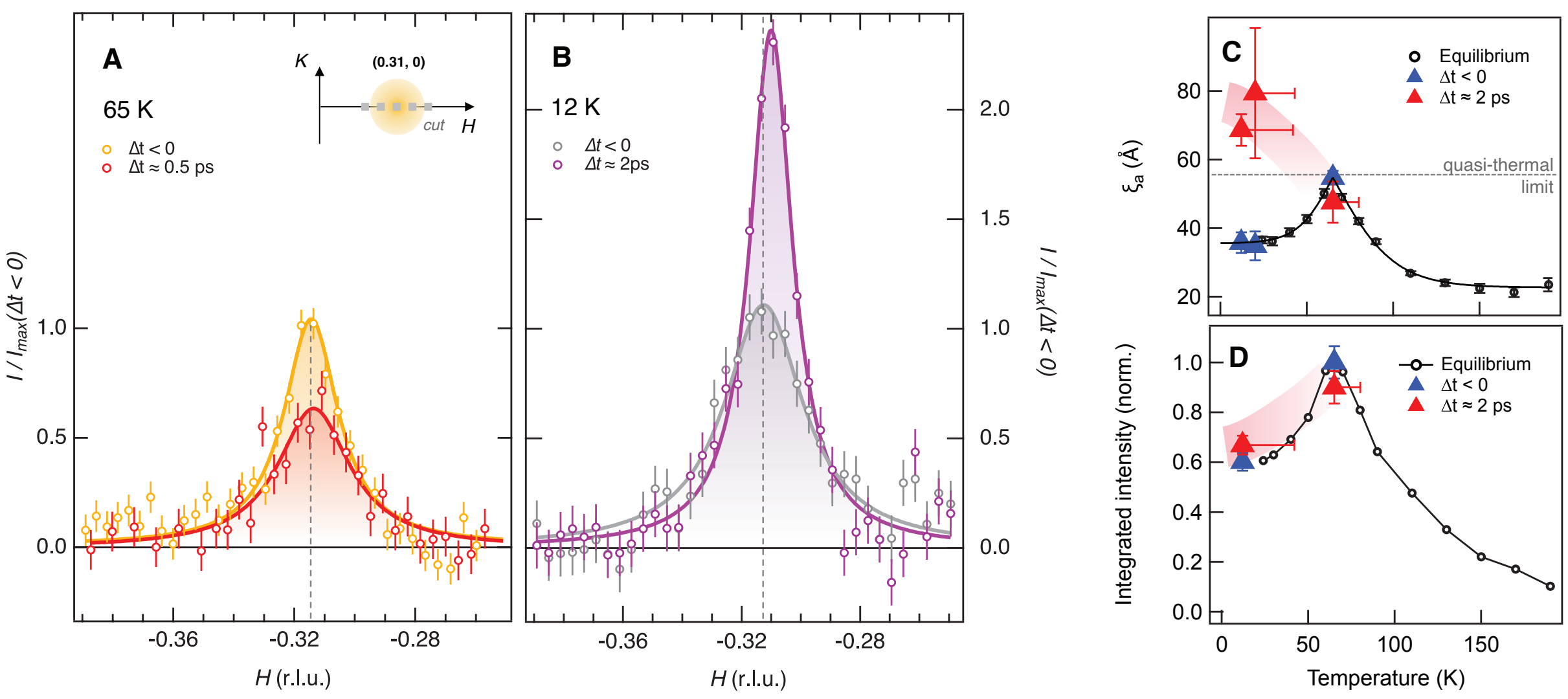

**Fig. 2. X-ray scattering profiles of CDW peak before and after photo-excitation. (A-B)** *H*-scan plots of CDW scattering signal, after background subtraction, in equilibrium (yellow symbols in **A**, and gray symbols in **B**) and following photo-excitation (red symbols in **A**, purple symbols in **B**) at 65K **(A)** and 12 K **(B)**. The pump fluence was 0.05 mJ/cm$^2$. Vertical dashed lines mark the wavevector of the peak at negative delays. Similar results to **(B)** were obtained using a different detector and at T = 20 K (Fig.S2) (*36*). Error bars are 1 SD. **(C-D)** Correlation length $\xi_a$ along the $a$-axis (**C**) and integrated intensity $I_{\text{int}}$ (**D**) as a function of temperature measured at a synchrotron (black squares), and at the LCLS at negative delays (blue triangles), and around ≈ 2 ps delay time (red triangles). For the data at 65 K the 2 ps values were extrapolated from 0.5 ps assuming the decay dynamics reported in Fig. 1C. Error bars are evaluated as 1 SD for the fit coefficient. The error bars on the temperature value of the pumped data reflect possible heating effects (*36*).

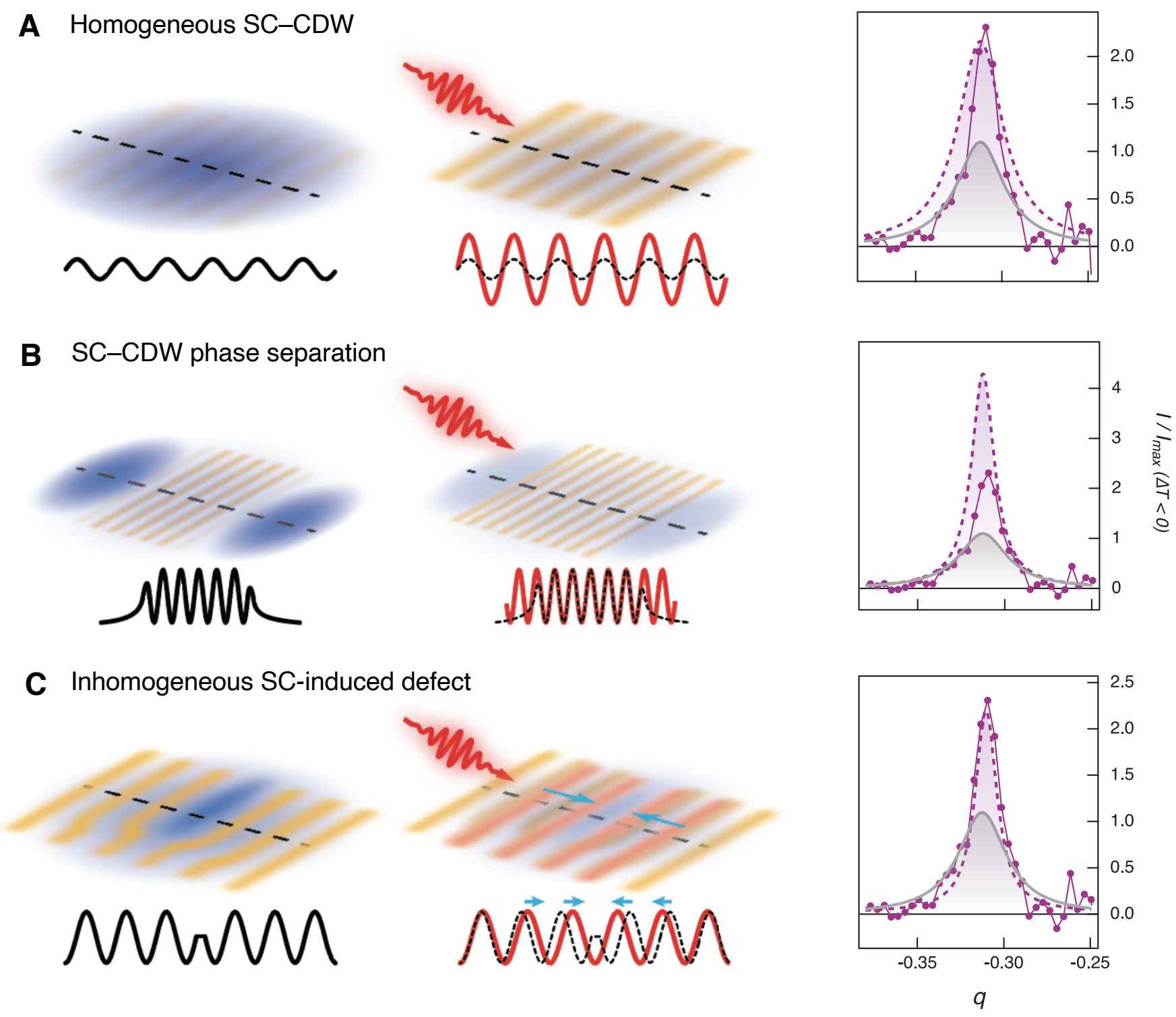

**Fig. 3. Scenarios for SC and CDW interactions.** (**A-B-C**) Models of the spatial arrangements of CDW and SC regions before and after photo-excitations. The first column represents the ground state in each case, and the middle column represents the state following the quench of the superconducting state by a laser pulse. The third column shows the simulated scattering profile for each of these cases for the equilibrium (gray solid lines) and photo-excited case (purple dashed lines) (*36*). The profiles are compared to the experimental data from Fig. 2B (purple solid lines and symbols). The three scenarios considered are: **(A)**, coexisting homogeneous SC and CDW orders, **(B),** mesoscale (>1 nm) phase separation between SC and CDW domains, and **(C),** SC-induced topological defect within the CDW domain (*36*). The data are well reproduced in this latter case sketched in **(C)**. Note: in (**C**) the SC state is inhomogeneous and developing more strongly around the defects, but still permeates the whole sample volume.

**References and Notes:**

1. J. C. S. Davis, D. H. Lee, Concepts relating magnetic interactions, intertwined electronic orders, and strongly correlated superconductivity. *PNAS*. **110**, 17623–17630 (2013).
2. E. Fradkin, S. A. Kivelson, J. M. Tranquada, *Colloquium*: Theory of intertwined orders in high temperature superconductors. *Rev. Mod. Phys.* **87**, 457–482 (2015).
3. J. M. Tranquada, B. J. Sternlieb, J. D. Axe, Y. Nakamura, S. Uchida, Evidence for stripe correlations of spins and holes in copper oxide superconductors. *Nature*. **375**, 561–563 (1995).
4. J. E. Hoffman, A Four Unit Cell Periodic Pattern of Quasi-Particle States Surrounding Vortex Cores in $Bi_2Sr_2CaCu_2O_{8+\delta}$. *Science*. **295**, 466–469 (2002).
5. C. Howald, H. Eisaki, N. Kaneko, A. Kapitulnik, Coexistence of periodic modulation of quasiparticle states and superconductivity in $Bi_2Sr_2CaCu_2O_{8+\delta}$. *PNAS*. **100**, 9705–9709 (2003).
6. M. Vershinin *et al.*, Local Ordering in the Pseudogap State of the High-Tc Superconductor $Bi_2Sr_2CaCu_2O_{8+\delta}$. *Science*. **303**, 1995–1998 (2004).
7. P. Abbamonte *et al.*, Spatially modulated “Mottness” in $La_{2-x}$ $Ba_xCuO_4$. *Nature Physics*. **1**, 155–158 (2005).
8. G. Ghiringhelli *et al.*, Long-Range Incommensurate Charge Fluctuations in $(Y,Nd)Ba_2Cu_3O_{6+x}$. *Science*. **337**, 821–825 (2012).
9. W. Tabis *et al.*, Charge order and its connection with Fermi-liquid charge transport in a pristine high-$T_c$ cuprate. *Nature Communications*. **5**, 1–6 (2014).
10. E. H. da Silva Neto *et al.*, Ubiquitous interplay between charge ordering and high-temperature superconductivity in cuprates. *Science*. **343**, 393–396 (2014).
11. G. Campi *et al.*, Inhomogeneity of charge-density-wave order and quenched disorder in a high-$T_c$ superconductor. *Nature*. **525**, 359–362 (2015).
12. A. Mesaros *et al.*, Commensurate $4a_0$-period charge density modulations throughout the $Bi_2Sr_2CaCu_2O_{8+x}$ pseudogap regime. *PNAS*. **113**, 12661–12666 (2016).
13. R. Comin, A. Damascelli, Resonant X-Ray Scattering Studies of Charge Order in Cuprates. *Annu. Rev. Condens. Matter Phys.* **7**, 369–405 (2016).
14. R. Arpaia *et al.*, Dynamical charge density fluctuations pervading the phase diagram of a Cu-based high-Tc superconductor. *Science*. **365**, 906–910 (2019).
15. J. Chang *et al.*, Direct observation of competition between superconductivity and charge density wave order in $YBa_2Cu_3O_{6.67}$. *Nature Physics*. **8**, 871–876 (2012).
16. A. Mesaros *et al.*, Topological Defects Coupling Smectic Modulations to Intra–Unit-Cell Nematicity in Cuprates. *Science*. **333**, 426–430 (2011).
17. T. Wu *et al.*, Magnetic-field-induced charge-stripe order in the high-temperature superconductor $YBa_2Cu_3Oy$. *Nature*. **477**, 191–194 (2011).

18. I. Vinograd *et al.*, Locally commensurate charge-density wave with three-unit-cell periodicity in YBa2Cu3Oy. *Nature Communications*. **12**, 1–11 (2021).

19. S. Gerber *et al.*, Three-dimensional charge density wave order in $YBa_2Cu_3O_{6.67}$ at high magnetic fields. *Science*. **350**, 949–952 (2015).

20. H. H. Kim *et al.*, Uniaxial pressure control of competing orders in a high-temperature superconductor. *Science*. **362**, 1040–1044 (2018).

21. D. Fausti *et al.*, Light-Induced Superconductivity in a Stripe-Ordered Cuprate. *Science*. **331**, 189–191 (2011).

22. G. Coslovich *et al.*, Competition Between the Pseudogap and Superconducting States of $Bi_2Sr_2Ca_{0.92}Y_{0.08}Cu_2O_{8+\delta}$ Single Crystals Revealed by Ultrafast Broadband Optical Reflectivity. *Phys. Rev. Lett.* **110**, 107003 (2013).

23. C. Giannetti *et al.*, Ultrafast optical spectroscopy of strongly correlated materials and high-temperature superconductors: a non-equilibrium approach. *Advances in Physics*. **65** (2016), pp. 58–238.

24. A. Kogar *et al.*, Light-induced charge density wave in $LaTe_3$. *Nature Physics*. **16**, 159–163 (2020).

25. W. S. Lee *et al.*, Phase fluctuations and the absence of topological defects in a photo-excited charge-ordered nickelate. *Nature Communications*. **3**, 838 (2012).

26. R. I. Tobey *et al.*, Evolution of three-dimensional correlations during the photoinduced melting of antiferromagnetic order in $La_{0.5}Sr_{1.5}MnO_4$. *Phys. Rev. B*. **86**, 064425 (2012).

27. P. Beaud *et al.*, A time-dependent order parameter for ultrafast photoinduced phase transitions. *Nat Mater*. **13**, 923–927 (2014).

28. M. Mitrano *et al.*, Ultrafast time-resolved x-ray scattering reveals diffusive charge order dynamics in $La_{2-x}Ba_xCuO_4$. *Sci. Adv.* **5**, eaax3346 (2019).

29. H. Jang *et al.*, Time-resolved resonant elastic soft x-ray scattering at Pohang Accelerator Laboratory X-ray Free Electron Laser. *Rev Sci Instrum*. **91**, 083904 (2020).

30. P. Kusar *et al.*, Controlled Vaporization of the Superconducting Condensate in Cuprate Superconductors by Femtosecond Photoexcitation. *Phys. Rev. Lett.* **101**, 227001 (2008).

31. G. Coslovich *et al.*, Evidence for a photoinduced nonthermal superconducting-to-normal-state phase transition in overdoped $Bi_2Sr_2Ca_{0.92}Y_{0.08}Cu_2O_{8+\delta}$. *Phys. Rev. B*. **83**, 064519 (2011).

32. L. Stojchevska *et al.*, Mechanisms of nonthermal destruction of the superconducting state and melting of the charge-density-wave state by femtosecond laser pulses. *Phys. Rev. B*. **84**, 180507 (2011).

33. Y. Toda *et al.*, Quasiparticle relaxation dynamics in underdoped $Bi_2Sr_2CaCu_2O_{8+\delta}$ by two-color pump-probe spectroscopy. *Phys. Rev. B*. **84**, 174516 (2011).

34. G. Coslovich *et al.*, Ultrafast dynamics of vibrational symmetry breaking in a charge-ordered nickelate. *Sci. Adv.* **3**, e1600735 (2017).

35. M. Mitrano *et al.*, Evidence for photoinduced sliding of the charge-order condensate in $La_{1.875}Ba_{0.125}CuO_4$. *Phys. Rev. B*. **100**, 205125 (2019).

36. See Supplementary Materials.

37. R. A. Kaindl, Ultrafast Mid-Infrared Response of $YBa_2Cu_3O_{7-\delta}$. *Science*. **287**, 470–473 (2000).

38. A. Zong *et al.*, Evidence for topological defects in a photoinduced phase transition. *Nature Physics*. **9**, 1 (2018).

39. Y. F. Kung *et al.*, Time-dependent charge-order and spin-order recovery in striped systems. *Phys. Rev. B*. **88**, 125114 (2013).

40. Y. Yu, S. A. Kivelson, Fragile superconductivity in the presence of weakly disordered charge density waves. *Phys. Rev. B*. **99**, 144513 (2019).

41. A. Moor, A. F. Volkov, K. B. Efetov, Topological defects in systems with two competing order parameters: Application to superconductors with charge- and spin-density waves. *Phys. Rev. B*. **90**, 762 (2014).

42. C. Chan, Interplay between pair density waves and random field disorders in the pseudogap regime of cuprate superconductors. *Phys. Rev. B*. **93**, 184514 (2016).

43. T. Mertelj *et al.*, Incoherent Topological Defect Recombination Dynamics in $TbTe_3$. *Phys. Rev. Lett.* **110**, 156401 (2013).

44. N. P. Ong, D. D. Duggan, C. B. Kalem, T. W. Jing, P. A. Lee, in *Charge Density Waves in Solids* (Springer, Berlin, Heidelberg, 1985), pp. 387–395.

45. A. Bruchhausen *et al.*, Photoinduced chain-oxygen ordering in detwinned $YBa_2Cu_3O_{6.7}$ single crystals studied by reflectance-anisotropy spectroscopy. *Phys. Rev. B*. **69**, 224508 (2004).

46. N. P. Ong, Z. Z. Wang, in *Nonlinearity in Condensed Matter* (Springer, Berlin, Heidelberg, 1987), pp. 350–360.

47. A. J. Achkar *et al.*, Impact of Quenched Oxygen Disorder on Charge Density Wave Order in $YBa_2Cu_3O_{6+x}$. *Phys. Rev. Lett.* **113**, 107002 (2014).

48. C. Chen, L. Su, A. H. C. Neto, V. M. Pereira, Discommensuration-driven superconductivity in the charge density wave phases of transition-metal dichalcogenides. *Phys. Rev. B*. **99**, 121108 (2019).

49. S. D. Edkins *et al.*, Magnetic field–induced pair density wave state in the cuprate vortex halo. *Science*. **364**, 976–980 (2019).

50. M. H. Hamidian *et al.*, Detection of a Cooper-pair density wave in $Bi_2Sr_2CaCu_2O_{8+x}$. *Nature*. **532**, 343–347 (2016).

51. M. Saint-Paul *et al.*, Ultrasound study on $YBa_2Cu_3O_{7-\delta}$ and $GdBa_2Cu_3O_{7-\delta}$ single crystals. *Solid State Communications*. **69**, 1161–1163 (1989).

52. S. Wandel *et al.*, Dataset for “Enhanced charge density wave coherence in a light-quenched high temperature superconductor,, doi:10.5281/zenodo.6020036.

53. R. Comin *et al.*, Broken translational and rotational symmetry via charge stripe order in underdoped $YBa_2Cu_3O_{6+y}$. *Science*. **347**, 1335–1339 (2015).

54. J. W. Loram, K. A. Mirza, J. R. Cooper, W. Y. Liang, Electronic specific heat of $YBa_2Cu_3O_{6+x}$ from 1.8 to 300 K. *Phys. Rev. Lett.* **71**, 1740–1743 (1993).

55. N. E. Phillips *et al.*, Specific heat of $YBa_2Cu_3O_7$. *Journal of Superconductivity*. **7**, 251–255 (1994).

56. J.-H. She, J. Zaanen, A. Bishop, A. Balatsky, Stability of quantum critical points in the presence of competing orders. *Phys. Rev. B*. **82**, 165128 (2010).

57. Z. Sun, A. J. Millis, Transient Trapping into Metastable States in Systems with Competing Orders. *Phys. Rev. X*. **10**, 021028 (2020).

58. W. L. McMillan, Theory of discommensurations and the commensurate-incommensurate charge-density-wave phase transition. *Phys. Rev. B*. **14**, 1496 (1976).

**Acknowledgements:** We thank Steven Kivelson, Mariano Trigo, Matteo Mitrano, Marta Zonno, Sydney Dufresne and Vincent Esposito for useful discussions.
**Funding:** This work, and the use of the Linac Coherent Light Source (LCLS), SLAC National Accelerator Laboratory, is supported by the U.S. Department of Energy (DOE), Office of Science, Office of Basic Energy Sciences (BES) under Contract No. DE-AC02-76SF00515. The SXR Instrument is funded by a consortium whose membership includes the LCLS, Stanford University through the Stanford Institute for Materials Energy Sciences (SIMES), Lawrence Berkeley National Laboratory (LBNL), University of Hamburg through the BMBF priority program FSP 301, and the Center for Free Electron Laser Science (CFEL). Part of the research described in this paper was performed at the Canadian Light Source, a national research facility of the University of Saskatchewan, which is supported by the Canada Foundation for Innovation (CFI), the Natural Sciences and Engineering Research Council (NSERC), the National Research Council (NRC), the Canadian Institutes of Health Research (CIHR), the Government of Saskatchewan, and the University of Saskatchewan. Use of the Stanford Synchrotron Radiation Lightsource (SSRL), SLAC National Accelerator Laboratory, is supported by the DOE BES under Contract No. DE-AC02-76SF00515. This research was undertaken thanks in part to funding from the Max Planck-UBC-UTokyo Centre for Quantum Materials and the Canada First Research Excellence Fund, Quantum Materials and Future Technologies Program. This project is funded in part by the Killam, Alfred P. Sloan, and Natural Sciences and Engineering Research Council of Canada’s (NSERC’s) Steacie Memorial Fellowships (A.D.); the Alexander von Humboldt Fellowship (A.D.); the Canada Research Chairs Program (A.D.); NSERC, Canada Foundation for Innovation (CFI); British Columbia Knowledge Development Fund (BCKDF); and CIFAR Quantum Materials Program. E.H.d.S.N. acknowledges prior support from the Max Planck-UBC postdoctoral fellowship, and current support from the Alfred P. Sloan Fellowship in Physics and the National Science Foundation under Grant No. 1845994 and No. 2034345. This work was partially supported by UC Davis start-up funds. J.J.T. acknowledges support from the DOE BES, Materials Sciences and Engineering Division, under Contract DE- AC02-76SF0051, through the Early Career Research Program. R.A.K. was supported by the DOE BES Materials Sciences and Engineering

Division under Contract No. DE-AC02-05-CH11231. A.F.K. was supported by NSF DMR-1752713.

**Author contributions:** G.C., J.J.T., C.G., F.B. and E.H.d.S.N. conceived and designed the experiments; G.C., F.B., E.H.d.S.N., J.J.T., S.W., J.D.K., G.L.D., A.H.R., M.P.M., S.P.M., W.F.S., M-F.L., and T.B. performed the ultrafast resonant soft x-ray scattering measurements; S.W., L.S., G.C., M.X.N., S.Z., Y.W., S.B.W. and M.H.S. analyzed the ultrafast resonant soft x-ray scattering data; F.B., E.H.d.S.N., F.H., R.S., J-S.L., D.G.H., G.C., performed synchrotron-based soft x-ray scattering measurements; G.C., S.W. and W.He. performed the transient optical reflectivity measurements; G.C. carried out the theoretical simulations with the assistance of A.F.K. and F.B.; R.L., D.B., W.Ha., grew and prepared the YBCO single crystals; G.C. wrote the manuscript with input from S.W., F.B., J.J.T., E.H.d.S.N., A.F.K., R.A.K., A.D. and contributions from all coauthors.

**Competing interests:** The authors declare no competing interests.

**Data and materials availability:** All experimental data shown in the main text and in the supplementary materials, as well as the simulation codes, are accessible at the Zenodo repository (*52*).

**Supplementary materials:**

Materials and methods

Figs. S1-S8

# Supplementary Materials for

## Enhanced charge density wave coherence in a light-quenched, high-temperature superconductor

**Authors:** S. Wandel, F. Boschini, E.H. da Silva Neto, L. Shen, M.X. Na, S. Zohar, Y.Wang, S.B. Welch, M.H. Seaberg, J.D. Koralek, G.L. Dakovski, W. Hettel, M-F. Lin, S.P. Moeller, W. F. Schlotter, A. H. Reid, M.P. Minitti, T. Boyle, F. He, R. Sutarto, R. Liang, D. Bonn, W. Hardy, R.A. Kaindl, D.G. Hawthorn, J-S. Lee, A.F. Kemper, A. Damascelli, C. Giannetti, J.J. Turner, G. Coslovich*

Correspondence to: gcoslovich@slac.stanford.edu.

**This PDF file includes:**

Materials and Methods
Figs. S1 to S8

## Materials and Methods

Equilibrium X-ray Scattering Characterization

Equilibrium resonant energy-integrated x-ray scattering (RXS) measurements were performed at the REIXS beamline of the Canadian Light Source (CLS) using a 4-circle diffractometer (operating in a low-$10^{-10}$ Torr ultra-high-vacuum chamber), with a photon flux of $\approx 5 \cdot 10^{12}$ photons/s and $\Delta E/E \approx 2 \cdot 10^{-4}$ energy resolution. *H*-scans (*H*,0,*L*) of the reciprocal space were performed by rocking the sample angle ($\theta$) at fixed detector position ($\theta_{det}$=170°), in a temperature range from 20 K to 250 K. Data were collected for incoming photon polarization either vertical (σ polarization) or horizontal (π polarization) to the scattering plane. The samples were pre-oriented using Laue diffraction and mounted with the a-axis in the scattering plane. The sample was then finely aligned in the RXS chamber via detection of the (0,0,2) and (1,0,2) Bragg peaks.

In an effort to maximize the charge-order signal, the incident photon energy was tuned resonant to the Cu-$L_{3/2}$ transition (≈ 932 eV, see the total fluorescence yield in Fig. S1A) and the multi-channel-plate detector was fixed at $\theta_{det}$=170° ($L \approx 1.45$ r.l.u. for $H \approx 0.31$ r.l.u.) (*8*, *13*, *53*).

The charge-order signal appears on the top of a fluorescence background (Fig. S1B). The RXS scans ($Q_{||}$,0,$L$) are well fit by a Lorentzian and a cubic background (*13*, *53*):

$$I_{RXS}\left(Q_{||}\right) = \frac{I_{\text{int}}}{\pi} \frac{\Delta Q_{||}}{\left(\frac{Q_{||}-Q_{CDW}}{\Delta Q_{||}}\right)^2+1} + \sum_{n=0}^{3} a_n {Q_{||}}^n, \tag{S1}$$

where $Q_{CDW}$ is the CDW-peak position, $I_{\text{int}}$ is the integrated intensity along the *H* profile and $\Delta Q_{||}$ the half-width-half-maximum. Fig. S1B displays the CDW peak at 60 K and the fluorescence background obtained at 300 K, fitted with a cubic polynomial. Figs. S1C-D show the peak for different orientations of the sample and at different temperatures after subtraction of the fluorescence background. In agreement with previous results on similar samples (*53*), we obtain almost identical longitudinal correlation lengths, i.e., the correlation length along $Q_{CDW}$, for the (0.31,0,1.45) and (0,0.31,1.45) peaks (Fig. 2C). The correlation length for the CLS data were corrected considering the angular acceptance of the MCP (≈ 3.8 deg). Panels A and B of Fig.3 in the main text show the longitudinal correlation-length ($\propto {\Delta Q_{||}}^{-1}$) and the integrated intensity of the CDW peak along the *H* profile (normalized to the value at T=65 K, extrapolated by a cubic spline) as a function of temperature.

Because of the stronger signal relative to the fluorescence background, the LCLS experiment was performed only for the *H*-scan (*H*,0,1.45). This scan was further crosschecked at the Stanford Synchrotron Radiation Lightsource (SSRL), BL 13-3, using similar experimental conditions and a 2D in-vacuum detector. The use of a 2D detector allowed to achieve better angular resolution than at CLS, and the resulting FWHM (≈ 0.022 r.l.u.) at 65K is in line with the resolution-corrected estimate from the CLS data.

Ultrafast Laser Pump - X-ray Scattering Probe Methods

Ultrafast energy-integrated RXS measurements were performed using the LCLS x-ray free-electron laser (FEL) at the SLAC National Accelerator Laboratory. The measurements reported were carried out at the Soft X-ray (SXR) beamline using the RSXS end station. The chamber pressure was maintained at ≈$10^{-9}$ mbar while the temperature was controlled using a helium-flow cryostat. Measurements reported above the critical temperature ($T_C$) were done at 65 K where the correlation length and the CDW intensity are maximal. Measurements below $T_c$ were done at 20 K where superconductivity and CDW coexist and their competition, i.e., the reduction of CDW amplitude, is maximum. Scattering measurements were performed in π-geometry using horizontal x-ray polarization. X-ray pulses were delivered to the beamline at a repetition rate of 120 Hz. The photon energy of the x-ray probe pulses was tuned to be resonant with the Cu-$L_3$ absorption edge (see the x-ray absorption spectrum in Fig. S1A), and the spectral bandwidth following the grating monochromator was 0.3 eV. The x-ray pulses were focused to a spot size of 100-150 μm in diameter, with average pulse duration of 100 fs and pulse energy of ≈ 1 μJ. X-ray fluence was adjusted to ensure that no x-ray pulse self-induced perturbation of the CDW peak was observed.

Optical pump pulses were generated using a Ti:sapphire chirped-pulse amplifier and compressed to ≈ 100 fs pulse duration, measured with a commercial SPIDER instrument. The 800-nm laser pulses were propagated collinearly with the x-ray pulses at a repetition rate of 120 Hz. The optical pulses were focused to a spot size of 400 μm and had a pulse duration of ≈ 100 fs. The laser fluence was controlled with a $\lambda/2$ waveplate and was varied from 20 – 200 μJ/cm$^2$, which is strong enough to quench superconductivity below $T_c$ but weak enough to minimize parasitic local heating effects. Spatial overlap between the optical pump pulses and x-ray probe pulses was performed using fluorescence generated from a frosted Ce:YAG crystal in the sample plane. Timing synchronization was established by monitoring reflectivity variation of a thin-film $Si_3N_4$ sample while scanning the time delay of the optical laser pulses. CDW dynamics were measured by monitoring the strength of the CDW diffraction peak while the relative time delay between the optical laser and x-rays was rastered across a 20-ps window using a motorized, encoded delay stage. Fine-timing corrections were achieved using the phase cavity after the undulator, which generates and distributes an RF reference signal to the experimental hutch. The time resolution was determined to be approximately 200 fs.

The CDW diffraction peak around wavevector $\boldsymbol{Q}$ = (0.31,0,1.45) was initially monitored using two avalanche photodiodes (APD) movable within the chamber. The sample angle was rocked about the CDW Bragg peak over a 25-degree range while the detector angle was held fixed ($\theta_{\text{det}}$=170°). Once $\boldsymbol{Q}_{\text{CDW}}$ was established, the sample angle was fixed and the CDW peak was pumped with the optical laser. Above $T_C$, photo-induced rocking curves were measured at a time delay corresponding to the largest CDW melting amplitude as determined by the CDW dynamical curves ($\tau \approx$ 0.5 ps). Below $T_C$, rocking curves were measured at a time delay corresponding to the largest CDW enhancement ($\Delta\tau \approx$ 2 ps). The negative delay signal was obtained by introducing an event sequencer of the optical laser in which every other shot was delayed by 10 ns. This delay provides ample time for the system to equilibrate, effectively producing a sequence of "pumped" and "unpumped" shots (negatively delayed). Shot-to-shot intensity fluctuations of the FEL were monitored using a multi-channel plate (MCP) detector directly after the monochromator. The MCP detector signal included a large-amplitude ringing component in the time domain, which was subsequently Fourier-transformed and frequency-filtered to eliminate low-frequency noise for $I_0$ normalization. The APD response was processed using a Savitzky-Golay filter with a second-order polynomial fit and by Singular Value Decomposition (SVD) methods. We obtained similar results with the two type of analysis.

To obtain better momentum resolution two strategies were adopted. The first consisted of placing a 1 mm aperture in front of the APD, leading to a resolution of 0.0015 r.l.u., but limited signal-to-noise due to the low signal. These early data are shown Fig. S2. A better solution consisted in the use of a 2D fiber-MCP detector that was commissioned for this experiment. The detail on the setup and performance of the detector will be described elsewhere. The detector had an angular acceptance of 3.8° and allowed to achieve high momentum resolution, while also improving the signal-to-noise ratio. This was achieved by the use of a photon counting algorithm and by averaging slices of the detector at slightly different 2θ angles.

The dynamics of the optical-pump x-ray-scattering probe measurements were modeled with exponential decay components. Above $T_C$, the data is well-described by a single exponential decay convoluted with a gaussian function to simulate the combined pump-probe time resolution of the experiment (≈ 200 fs). The dynamics represents the initial CDW melting and the subsequent relaxation. Below $T_C$, the interplay of CDW with superconductivity creates an additional dynamic response characterized by the slow enhancement of CDW following the suppression of superconductivity. We represent this data with a sum of two exponential decays. The data above and below $T_C$ at various fluences are shows in Fig. S3 together with the dynamical fits.

To estimate the x-ray penetration depth we used the x-ray linear absorption value obtained from a fluorescence (TFY) measurement across the $L_2$ and $L_3$ edges. The low and high energy sides of the absorption scan have been rescaled using the database values available online from the Center for X-ray Optics (CXRO). Considering the experiment scattering geometry we can calculate a linear attenuation coefficient of $7 \times 10^4$ cm$^{-1}$ for the incident x-rays (average of E//a and E//c for $\theta$ = 45°) and $8 \times 10^4$ cm$^{-1}$ for the scattered x-rays. The resulting estimate of the x-ray penetration depth is $\ell_{\text{x-ray}}$=52 nm. To estimate the optical penetration depth of the pump beam we used the reported value from Ref. (*32*), of 85 nm.

Optical Pump-Optical Probe Measurements

In order to reveal the superconducting dynamic and its fluence dependence on the YBCO sample, optical pump-optical probe measurements were performed. Here 1.5 eV ultrashort pulses are used to photo-excite the sample at varying fluences, while a – weaker – fraction of the same pulse is used to observe the dynamical response as a function of time delay. The sample was mounted in a helium-flow cryostat and photo-excited at a pulse repetition rate of 120 Hz. The pump fluence was varied from 10 – 150 μJ/cm$^2$, while the probe fluence was fixed at about 1 μJ/cm$^2$. The detector measured the reflectivity of the sample as the time delay between the pump and probe pulses varied over a range of 12 ps using a motorized encoded stage. We observe a saturation of dynamical signal in the superconducting phase around 50 μJ/cm$^2$, as shown in Fig. S4A, indicating the occurrence of a photo-induced phased transition, in agreement with previous reports (*32*). The signal disappears above $T_C$, establishing its connection to the well-known Cooper pair relaxation dynamics typical of high-temperature superconductors (*23*). To prove this point we measured the response using a fluence above saturation (≈ 100 μJ/cm$^2$), illustrated in Fig. S4B. Here the saturated superconducting signal still dominates over other linear components, such as the lattice, normal state electrons or the pseudogap responses, which are equally present at 65K, while the superconducting component is heavily suppressed. To further establish the connection between the optical data measured at 800 nm and the superconducting condensate dynamics, we show a direct comparison to the transient MIR spectroscopy data measured resonantly to the superconducting gap on an analogous sample (*37*) (Fig. S5).

Fluence calibration

During each pump-probe measurement the fluence was estimated based on the pump energy per pulse and its diameter (FWHM) at the interaction point. The pump energy was monitored with a power meter before beam transport to the interaction point, and the beam transport losses were calibrated once for each measurement session. While relative fluence changes within each session are well calibrated by monitoring the relative pump energy variations, care needs to be taken when comparing absolute fluences between different measurement sessions, such as optical and FEL probe experiments.

For FEL measurements the uncertainty on the absolute fluence value is mostly due to two factors: i) the transport losses may vary slightly due to the alignment through the in-coupling mirror, which features a 2 mm-hole to allow the x-ray beam through (see Fig.1A scheme); we evaluated this uncertainty to be about 15%; ii) the beam diameter evaluation has a 10% uncertainty. This leads to about ≈ 35% uncertainty on the absolute fluence value. Since in the main text we consider data coming from different measurements sessions, we approximated the fluence values considering this uncertainty. For example, the fluence value of 0.05 mJ/cm$^2$ represents measurements in which the estimated fluence was in the range (50 ± 18) μJ/cm$^2$. We stress that we observed no substantial differences – within the noise of the measurements – in the FEL results within this fluence range. This is due to the saturated response of the superconducting signal at this fluence level (Fig. 1D).

When comparing FEL and optical pump-probe results, we considered the energy density deposited in the probed volume in each type of measurement. X-ray and optical probes have similar penetration depths, 52 nm and 42.5 nm respectively. Given that this small difference is within the uncertainty on the absolute fluence value, we compared data with the same fluence for simplicity. Our conclusions and main results remain unchanged even when applying this small correction for the comparison between x-ray and optical data.

Heating calculation

To estimate the pump-induced heating effect we assume the pump energy density to be fully absorbed by the sample within the penetration depth, neglecting diffusion effects. Inclusion of diffusion effects may lower this estimate, so we consider this as an upper bound. The energy density was estimated based on the fluence and penetration depth of the pump pulse. For a fluence of (50 ± 18) μJ/cm$^2$ the energy density is (5.8 ± 2.0) J/cm$^3$. We then used the temperature dependent value of the specific heat in Ref. (*54*) to estimate the temperature increase. The specific heat value was cross-checked with a separate study (*55*) and coincided within few %. For 0.05 mJ/cm$^2$ the estimated temperature increase value is (23 ± 5) K when the initial temperature is 20 K. The final maximum temperature in this case would be (43 ± 5) K. This temperature range estimate is reflected in the temperature error bars in Fig. 2C-D of the main text.

The heating dynamics shown in Fig. 1E is derived by combining the above lattice temperature estimation with an interpolation of the temperature dependent scattering peak strength shown in Fig. S1. To estimate the heating timescale we consider that, while a small subset of phonons may react on fast timescales (≈ 100 fs), the bulk of the lattice reacts to 1.5 eV laser excitations on a timescale of ≈ 2ps (*23*). This dynamics can be reproduced for positive delay times $\Delta t$ with an exponentially rising function of the form $1\text{-}e^{-\Delta t/\tau}$ with $\tau$ = 2 ps.

Time-dependent Ginzburg-Landau

The temporal evolution of the amplitude of the CDW and superconductivity (SC) order parameters can be modeled by a two-component time-dependent Ginzburg-Landau (TDGL) expansion of the free energy. The lowest-order symmetry-allowed interaction terms incorporating CDW and superconductivity (*2*, *56*) are: $F = F_{SC} + F_{CDW} + F_{\text{int}}$, where $F_{SC}$ and $F_{CDW}$ are the independent GL expansions up to the quartic terms for superconductivity and CDW, respectively: $F_k = \alpha_k|\psi_k|^2 + \beta_k|\psi_k|^4$. $\alpha_k$ and $\beta_k$ represent expansion coefficients for the complex order parameters that break different symmetries, $\psi_k$. $F_{\text{int}}$ contains the interaction term, $W|\psi_{SC}|^2|\psi_{CDW}|^2$, where $W$ determines the type and strength of interaction. A positive $W$ indicates competing orders, and if its value exceeds $2\sqrt{\beta_{SC}\beta_{CDW}}$ the weaker order parameter vanishes (*56*). Order coexistence is thus only allowed for $\frac{W}{2\sqrt{\beta_{SC}\beta_{CDW}}} < 1$. In the simplest case of homogeneous orders the relaxation dynamics for the amplitude can be represented as (*23*, *27*):

$$\frac{d^2\psi_k}{dt^2} + \alpha_k\omega_{0,k}\frac{d\psi_k}{dt} = -\omega_{0,k}^2\frac{dF}{d\psi_k^*}, \qquad \text{(S2)}$$

where $\alpha_k$ is the damping coefficient and $\omega_{0,k}$ is the frequency of the mode for each order. In our data, we do not observe coherent oscillations, within our time resolution of ≈ 200 fs, hence we can consider the highly overdamped regime, where $\alpha_{CDW} \gg 1$. In this case the second order term can be neglected and the dynamics is governed by the dynamical coefficient $\gamma_k = \frac{\omega_{0,k}}{\alpha_k}$.

The TDGL equations generally describe well the dynamics ensuing photo-induced melting of CDW (*23*, *27*, *39*), mostly dominated by the relaxation of the CDW amplitude. In the case of homogeneous and competing orders, TDGL predicts an homogeneous reaction of the order amplitude in response to the quench of the competitive counterpart (*22*, *57*). Because of the ps-scale relaxation dynamics observed at 65 K (Fig. 1C) TDGL predicts a picosecond homogeneous enhancement of the CDW amplitude for our experiment. However, the observed dynamics is dominated by the correlation length enhancement, with a smaller contribution derived from amplitude variations. A TDGL theory considering homogeneous and competitive orders, as in scenario A described in the main text, is therefore not appropriate to describe our data. A dynamical model considering the CDW phase dynamics, caused by the relaxation of the SC-induced CDW dislocations described in the main text, will be subject of future studies.

To further support the claim that scenario A is incompatible with the data we consider two extensions of the simple homogeneous TDGL treatment discussed so far:

1) Here we consider and exclude the possibility that inhomogeneity is light-induced. Such possibility is discussed in Ref. (*57*), where they considered the light-induced amplification of spatial fluctuations after the quench of two competing order parameters. This dynamics result into the formation of a metastable state, which we clearly do not observe in the time-resolved data, which exhibit relaxation on the order of single digit picosecond timescale. Moreover, as pointed out in the theory work by Sun et al., when only one of the interacting orders is quenched – as in our case – spatial fluctuations become irrelevant and the order parameter dynamics is expected to remain homogeneous.

2) We can also exclude a phenomenological quasi-thermal version of scenario A, where the order parameter dynamics follow an effective temperature model rather than the TDGL. This would be the case for order parameters with remarkably fast relaxation rates. As pointed out in the main text, the ultrafast CDW scattering response exhibit a strongly non-thermal behavior, exemplified in Fig. 3C and 3D. As the CDW response deviates significantly from an effective temperature, this model for scenario A can also be excluded.

Simulation of Resonant X-Ray Scattering Signals

We simulate the x-ray scattering signal by considering a distribution of CDW domains on a lattice with a CDW order parameter defined as

$$\psi_{\mathrm{CDW}}(r) = \psi_0(r) e^{i\boldsymbol{Q}_{\mathrm{CDW}}\cdot\mathbf{r}+\theta(r)}, \tag{S3}$$

where $\psi_0(r)$ and $\theta(r)$ are the CDW amplitude and phase across the sample. The charge distribution is $\rho(r) = \mathrm{Re}\,\psi_{\mathrm{CDW}}(r)$. The x-ray scattering signal for a given solid angle around $\boldsymbol{q}$ is proportional to the scattering cross-section

$$\frac{d\sigma}{d\Omega} = A(\boldsymbol{q})A^*(\boldsymbol{q}) = \left| r_0 \int_{Crystal} d\boldsymbol{r}\, \rho(\boldsymbol{r}) e^{-i\boldsymbol{q}\cdot\boldsymbol{r}} [\boldsymbol{\epsilon}\cdot\boldsymbol{\epsilon}'] \right|^2, \tag{S4}$$

where $A$ is the scattering amplitude for a transferred wavevector $\boldsymbol{q}$, and the right hand side is the squared amplitude of the Fourier Transform of the charge density modulations integrated over the whole crystal and multiplied by the polarization factor $[\boldsymbol{\epsilon}\cdot\boldsymbol{\epsilon}']$. The charge distribution for the crystal is considered to be neutral, and zero everywhere, except for each CDW domain adding a local charge modulation $\rho(r)$. The numerical simulation is performed considering 2000x2000 lattice sites (unit cells) in 2D, with an incommensurate CDW periodicity to reproduce the experimental $\boldsymbol{Q}_{\mathrm{CDW}}$ = (0.31,0) r.l.u.. To satisfy these conditions and considering Eq. (S3), the numerically computed charge distribution within a domain takes the form

$$\rho(x,y) = \psi_0(x,y) * \cos\left(\frac{2\pi}{a_{CDW}} * x + \theta_{\mathrm{Domain}} + \theta_{\mathrm{Noise}}(x,y)\right), \tag{S5}$$

where $\psi_0(x,y)$ is the amplitude of the CDW, $a_{CDW}$ is the CDW periodicity, $\theta_{\mathrm{Domain}}$ is a constant phase factor different for each domain, and $\theta_{\mathrm{Noise}}(x,y)$ is a "noise" phase factor varying site by site to simulate thermal fluctuations across the crystal.

The shape of the domains is generally elliptical with an exponentially decaying tail at the end of the domain to minimize spurious diffraction signals arising from the choice of domain shape. A filling factor of 20% is considered for simplicity in order to avoid overlaps between the randomly distributed CDW domains. Simulations at different filling factors, as well as larger system sizes, yield similar results with just a rescaled scattering intensity. The spatial map of a typical simulation is shown in Fig. S6A, with the corresponding distribution of the domain sizes shown in Fig. S6B. The distribution of CDW domain sizes is Gaussian-like and truncated at zero to avoid non-physical conditions. The width of the distribution is optimized a) to provide the best agreement with the experimental data, and b) to be in line with micro X-ray diffraction imaging data on a Hg-based cuprate displaying a similar CDW (*11*). While the nanoscale picture of the CDW domains is more complex and characterized by multiple types of defects (cfr. (*12*, *16*)), we consider this model as a reasonable simplification of the static CDW landscape, while we focus on superconductivity-induced variations.

The 2D scattering signal is obtained via FFT (see Fig. S6C) and is then convolved with a Gaussian function representing the experimental resolution. The projection of the peak along the H-direction in reciprocal space is then considered (see Fig. S6D). Finally, to remove any residual feature arising from speckle patterns and variations inherent to an individual simulation, the curves are fitted with a Lorentzian function, which can be directly compared to the data in the main text. In the example presented in Fig. S6, we see that the 65 K equilibrium experimental correlation length and lineshape are reproduced by considering a distribution of domains with an average radial size of ≈ 35 unit cells and periodicity $a_{CDW} = 3.225$ unit cells.

Using this approach, we simulated and investigated four representative cases, some of which are discussed in the main text and displayed in Fig.3 of the main text. In each case we repeated two identical simulations, one in "equilibrium" conditions, and then in the "pumped" state, with only control parameter being changed. In the case of defects, the properties of the CDW domain remained unaltered, with the only difference being the introduction of defects.

1) Homogenous CDW-SC: We first consider the case of two coexisting, yet competing, homogeneous CDW and SC orders, as described in the previous section. TDGL theory predicts an increase of amplitude of the CDW order after the quench of the SC state. To reproduce the fact that the CDW peak maximum signal is almost doubled after photo-excitation, we considered a fairly large (40%) increase of the CDW amplitude $\psi_0$ (Fig. S7A). As shown in Fig. S7A, the correlation length remains unchanged in this scenario, which is in disagreement with the experiment.

2) Expanding CDW domain: In this scenario, indicative of phase-separated SC and CDW domains (scenario B in main text, illustrated in Fig.3B), we consider a situation where the correlation length increases by virtue of expansion of the amplitude envelope $\psi_0(x, y)$ of the CDW domains. Such increase of domain size, results in an increase of filling factor and thus integrated intensity $I_{\text{int}}$. In particular, we considered two cases: a bidimensional expansion of the CDW domains, and a unidimensional expansion. The results of the simulation for the latter case, where the domain size almost double in one direction, are shown in Fig.S7B. In this scenario the increase of the peak strength is much larger than what experimentally observed by at least a factor two. This is because the integrated intensity scales linearly with the correlation length increase, $I_{\text{int}} \propto \xi_a$ . The case of bidimensional expansion produces an even larger discrepancy with respect to the experimental data, due to the quadratic scaling $I_{\text{int}} \propto {\xi_a}^2$ . Notably, this scenario also implies that we can exclude the presence of new large domains nucleating from previously SC regions.

3) Non-topological defects, vacancies: Here we consider non-topological defects, such as vacancies. This type of defects could be caused by quenched disorder, such as oxygen chain disorder, and may be reinforced by the formation of the SC in and around the defect. We remark that the laser pulse, especially at such low fluence (≈ 50 μJ/cm$^2$), should not affect the defect directly, but just the SC state that may be forming around it. In fact, an increase of correlation length, potentially indicative of defect annealing, is not observed above $T_C$, at 65K, where instead the opposite behavior is observed.

   We first considered the case of small vacancies, of the size of a unit cell. The simulated x-ray scattering signal for the case of ≈ 100 vacancies randomly distributed across the CDW domain and removed after photo-excitation is shown in Fig. S7C, and display a marked increase of integrated intensity. The correlation length (peak width) remains mostly unchanged. Different simulations with larger vacancies and a sparser distribution yielded similar results. The case of vacancies can be extended to other non-topological defects, which do not affect the CDW phase registry beyond a small local distortion. (Note: for more extended phase distortions the treatment of dislocations may be more appropriate, see next section). We thus rule out a scenario where the SC state induces non-topological defects within the CDW domains.

4) Topological defects, dislocations & discommensurations: We are thus left with the scenario of SC inducing topological defects, affecting the coherence of the CDW order. Within this class of defects, we consider dislocations and discommensurations, whose presence is well documented in cuprates (*12*, *16*, *18*). Their proliferation in the case of incommensurate order is related to the fact that phase modulations cost a vanishingly small amount of energy, compared to the finite energy cost of amplitude modulations. In this work, we consider such defects as being intrinsically stabilized by superconductivity, in order to explain the rapid enhancement of the CDW correlation length following the light-quench of superconductivity. We reiterate that such effect is only observed below $T_C$ and shows an ultrafast dynamics linked to the SC quench and recovery. In addition, we remark that the presence of topological defects, such as dislocations, tends to decrease the CDW correlation length (*25*, *38*, *39*). Therefore, the annihilation of topological defects is a legitimate candidate for explaining our experimental evidence. In the following, we show that discommensuration lines best reproduce our experimental data.

To simulate dislocations and discommensurations we consider additional amplitude and phase modulation factors in the CDW pattern within a domain (cfr. Eq. (S5)):

$$\rho(x,y) = \psi_0(x,y) * \cos\left(\frac{2\pi}{a_{CDW}} * x + \theta_{\mathrm{Domain}} + \theta_{\mathrm{Noise}}(x,y) + \theta_{\mathrm{Defect}}(x,y)\right), \qquad \text{(S6)}$$

where $\psi_0(x,y)$ and $\theta_{\mathrm{Defect}}(x,y)$ are respectively the amplitude and phase solutions in the presence of the defect. This approach is similar to previous studies (*12*, *18*). In the case of a dislocation we consider a CDW vortex solution using the following amplitude and phase factors in (S6):

$$\begin{aligned} \psi_0(x,y) &= 1 - \exp\left(-\frac{1}{2}\frac{(x-x_0)^2 + (y-y_0)^2}{\zeta^2}\right) \\ \theta_{\mathrm{Defect}}(x,y) &= \operatorname{atan}\left(\frac{y-y_0}{x-x_0}\right), \end{aligned} \qquad \text{(S7)}$$

where $(x_0, y_0)$ are the defect coordinates and $\zeta$ represents the width of the dislocation core. This dislocation represents a 2π vortex and, as previously reported in the case of Bi2212 (*16*), it generally appears in pairs with opposite winding phases. The motion of such dislocations can lead to their annihilation. This situation is schematically represented in Fig. S8A. In our simulation we consider an equilibrium condition with one dislocation per CDW domain, randomly located within the domain and with equal probability of having a +2π or -2π winding phase. In the pumped case all dislocations are removed (annihilated) and the original CDW pattern is restored. As shown in Fig.S8A the results of the simulation, display a similar increase (≈ 40%) for both the peak integrated intensity and its correlation length. This is mostly related to the fact that dislocations affect the correlation length along both planar directions, *a* and *b*, while in our data we observe a large variation of the longitudinal correlation length, i.e., along $a$ for the $\boldsymbol{Q}_{\text{CDW}}$ = (0.31,0) peak. Moreover, because of the balanced distribution of winding phases, this simulation does not show any momentum shift. We conclude that dislocations alone cannot reproduce the data reported in the main text.

We next consider the case of discommensurations (DCs), a particular type of defect associated with nearly commensurate CDWs, which was originally discussed by McMillan (*58*) and are broadly reported in transition metal dichalcogenides (*48*). DCs can be represented as sudden jumps in the CDW phase argument that can allow for a local commensurate pattern even for a globally incommensurate order. In our case we can represent DCs as 1D jumps in the phase factor $\theta_{\text{Defect}}(x, y)$ along the $\boldsymbol{Q}_{\text{CDW}}$ direction. They can be associated with a 1D core with vanishing CDW amplitude (*12*). For our simulations we considered several versions of DCs, with phase jumps of $\pm\pi$, $\pm\pi/2$ and $\pm 2\pi/3$, following the existing STS and NMR literature on the topic (*12*, *18*). The type that best represented the data is shown in Fig. S8B, as well as in Fig. S3C. We considered a core of about one unit cell, and a phase jump that decreases exponentially when moving away from the DC core. This is because of the boundary condition set at the edge of the domain, which we consider here as determined by static disorder, not affected within the ultrafast timescale of our experiment.

The simulation of the x-ray scattering signal for the case of DCs generally reproduces the core features of our experiment: large variation of correlation length along the *H* direction, together with a small variation of integrated intensity and a momentum shift. The increase in correlation length and integrated intensity are consistent with naive expectations after a single DC defect is removed from each domain. However, in this model the sign and magnitude of the shift in momentum are less obviously tied to the phase jump in the DC because a typical domain size is only ~6 periods of the CDW and hence the precise spatial profile of the phase jump matters for the resulting shift in Fourier amplitudes. In particular, we find very good agreements with the data when considering DCs whose spatial profile is a jump of -π with an exponential suppression towards the domain edges, as described above. This simulation is shown in Fig. S8B and reproduces the dramatic (≈ 95%) increase of correlation length, in combination with the small (≈ 10%) increase of peak amplitude (area) following photo-excitation and momentum shift of ≈ -.0022. The resulting Lorentzian fit is shown in Fig.3C of the main text. This result suggests that the presence of superconductivity induces DC lines resulting in local CDW arrangements with higher momenta (denser CDW lines) than the native CDW order. Such behavior will be further investigated in future studies.

Our data are reproduced considering DCs defects with an average density of 1 defect per domain. The domain size is of the order of the correlation length after photo-excitation, ≈ 7 nm. In this model, the average distance between each DC and then next defect is about 3-4 nm, as we expect more defects to be proliferate toward the boundary of the domains as expected from energy considerations (*42*). Defects present nearby the boundaries, including DCs, may be less visible in our experiment, as they produce smaller effects on the correlation length. In conclusion, we consider 3-4 nm to be a lower bound for the average spacing between DCs. Finally, we should also remark that, while dislocations alone cannot explain our data, we still expect them to be part of the superconductivity-induced inhomogeneous CDW landscape, as depicted in Fig. 3C. This is because multiple DC lines may converge or originate from the 2π phase vortex at the dislocation (*12*, *58*).

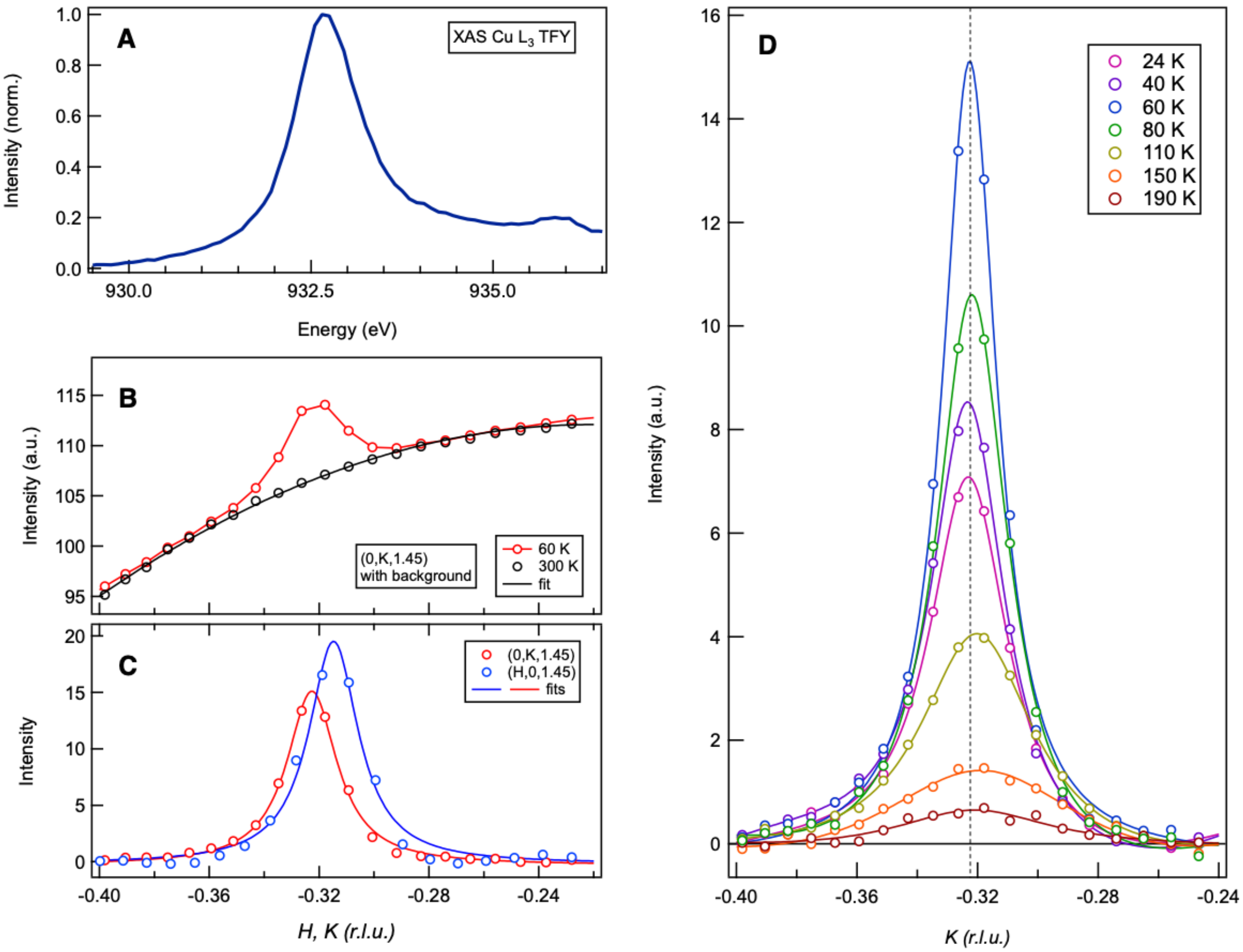

**Fig. S1. A,** X-ray absorption scan of YBCO around the Cu-$L_3$ edge measured at CLS as total electron yield (TEY) and total fluorescence yield (TFY). **(B-D)** K- and H-scans of the x-ray scattering signal on YBCO at various temperatures at the photon energy of 931.5 eV, collected at the Canadian Light Source (CLS). Panel **B** includes the fluorescence background, fitted with a polynomial function as described in the text, while in panels **C, D** the fluorescence background has been subtracted and the signal is fitted with a Lorentzian function.

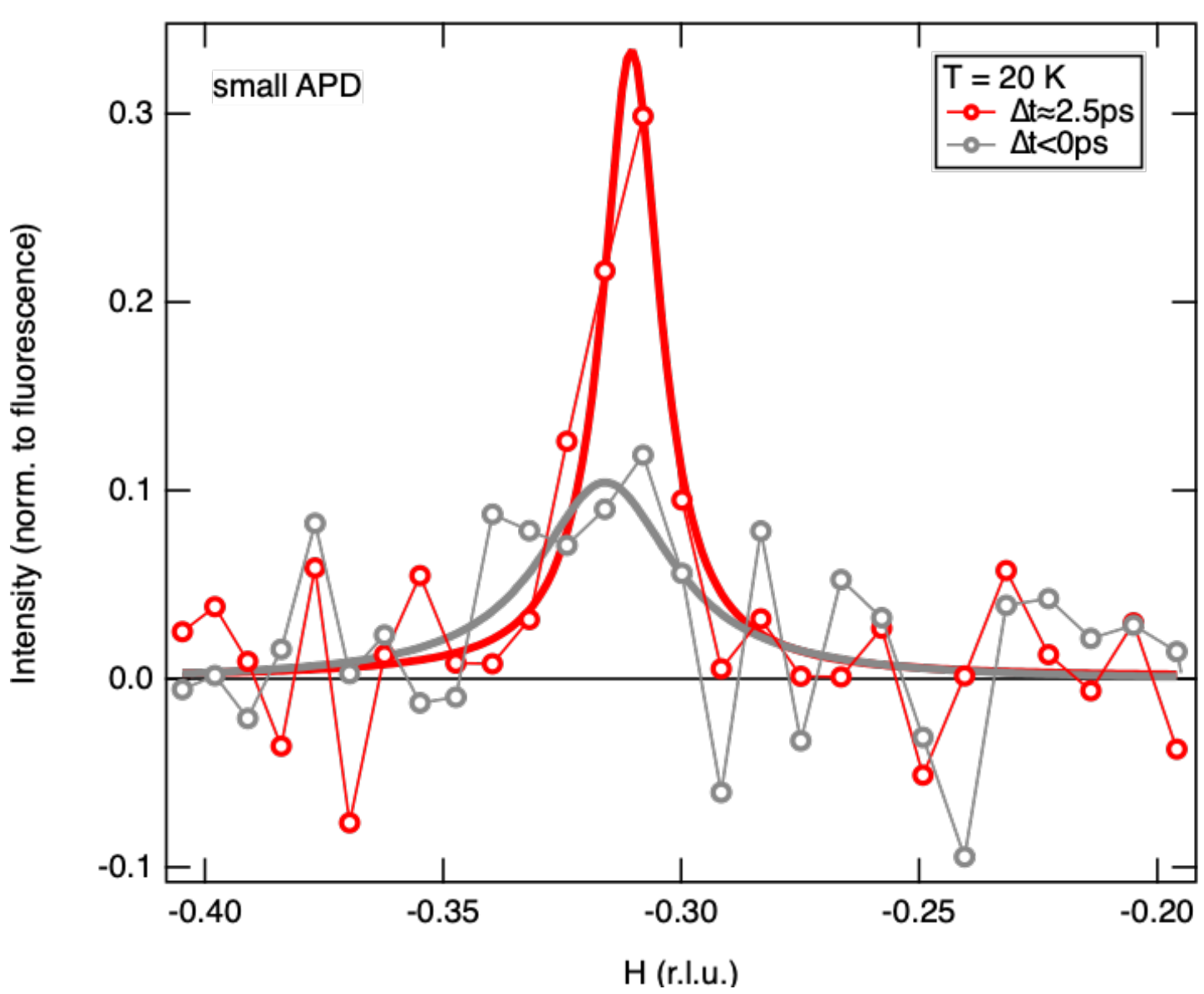

**Fig. S2.** Ultrafast RXS *H*-scan performed at T=20K and pump fluence ≈ 50 μJ/cm$^2$ on YBCO. The detector used in this case is an APD with a small aperture in front to improve the angular resolution. The estimated angular resolution was 0.0015 r.l.u. In this set of data, we obtain a relative increase of the CDW integrated intensity $\boldsymbol{I}_{\mathbf{int}}$ of ≈ 34%, compared to an increase in correlation length of ≈ 135%, from ≈ 35 Å to ≈ 81 Å. These correlation length data are considered in Fig.2C and they agree within the error bars with the data collected at 12 K. In the comparison we note the larger error bars for this data set, obtained with the APD. Due to the absence of a comparable curve at 65 K with this detector it was not possible to add the normalized integrated signal in Fig. 2D from this data set.

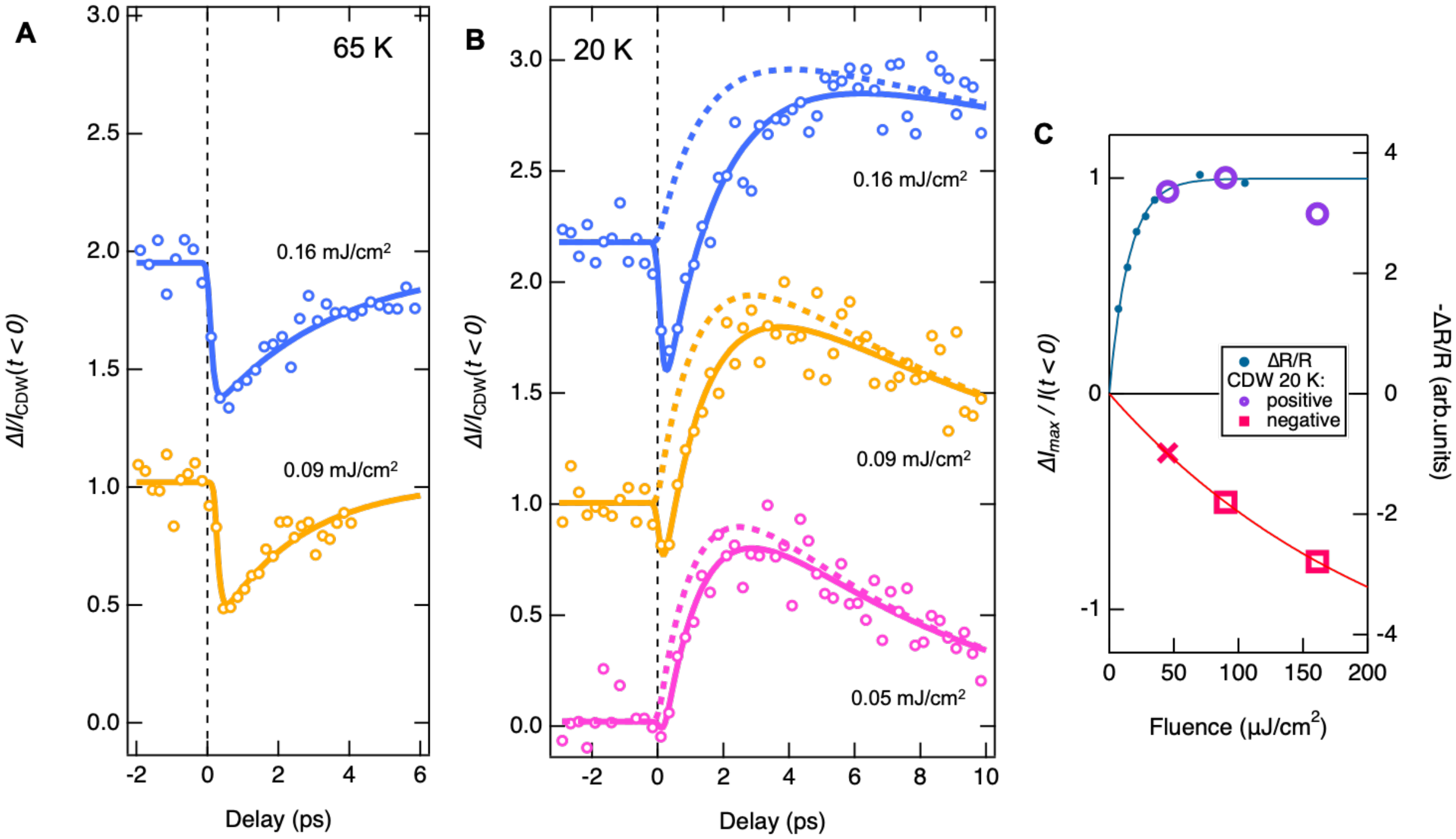

**Fig. S3. A,** Melting of the CDW for T=65 K, measured with a large APD capturing the signal around the peak of the CDW signal. **B,** Temporal evolution of the CDW peak signal at T=20 K. Solid lines represent a double exponential dynamics model considering a melting and an enhancement contribution. The dashed line in panel B isolates the enhancement contribution. These contributions are plotted in Fig.1E of the main text. **C** Fluence dependence of the CDW dynamical components, positive (circle) and negative (squares), at 20 K obtained from the fits of panel B and normalized to the intensity of the CDW peak before photo-excitation. The negative component data at 0.05 $mJ/cm^2$ is very small and its intensity was extrapolated from the fit of the fluence dependence (solid line). The fluence dependence of the optical data (from Fig. S4) is also shown (solid circles).

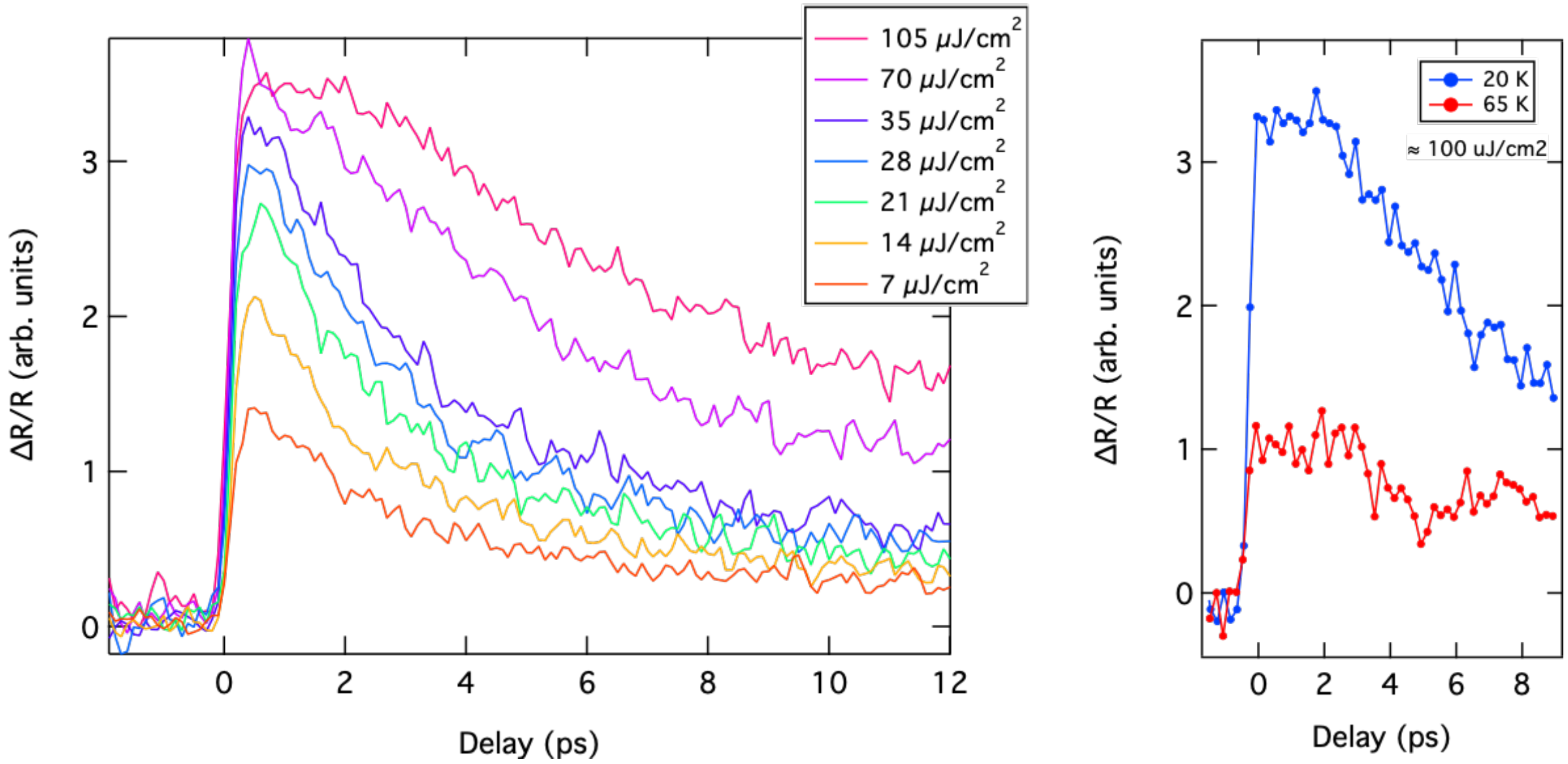

**Fig. S4. A,** Fluence dependence of the time-resolved reflectivity signal, 800 nm probe, in the superconducting phase of YBCO (T = 20 K). **B,** Time-resolved reflectivity signal at 20 K and 65 K for ≈ 100 µJ/cm$^2$. The signal drops dramatically around $T_C$ indicating the clear connection to the SC order.

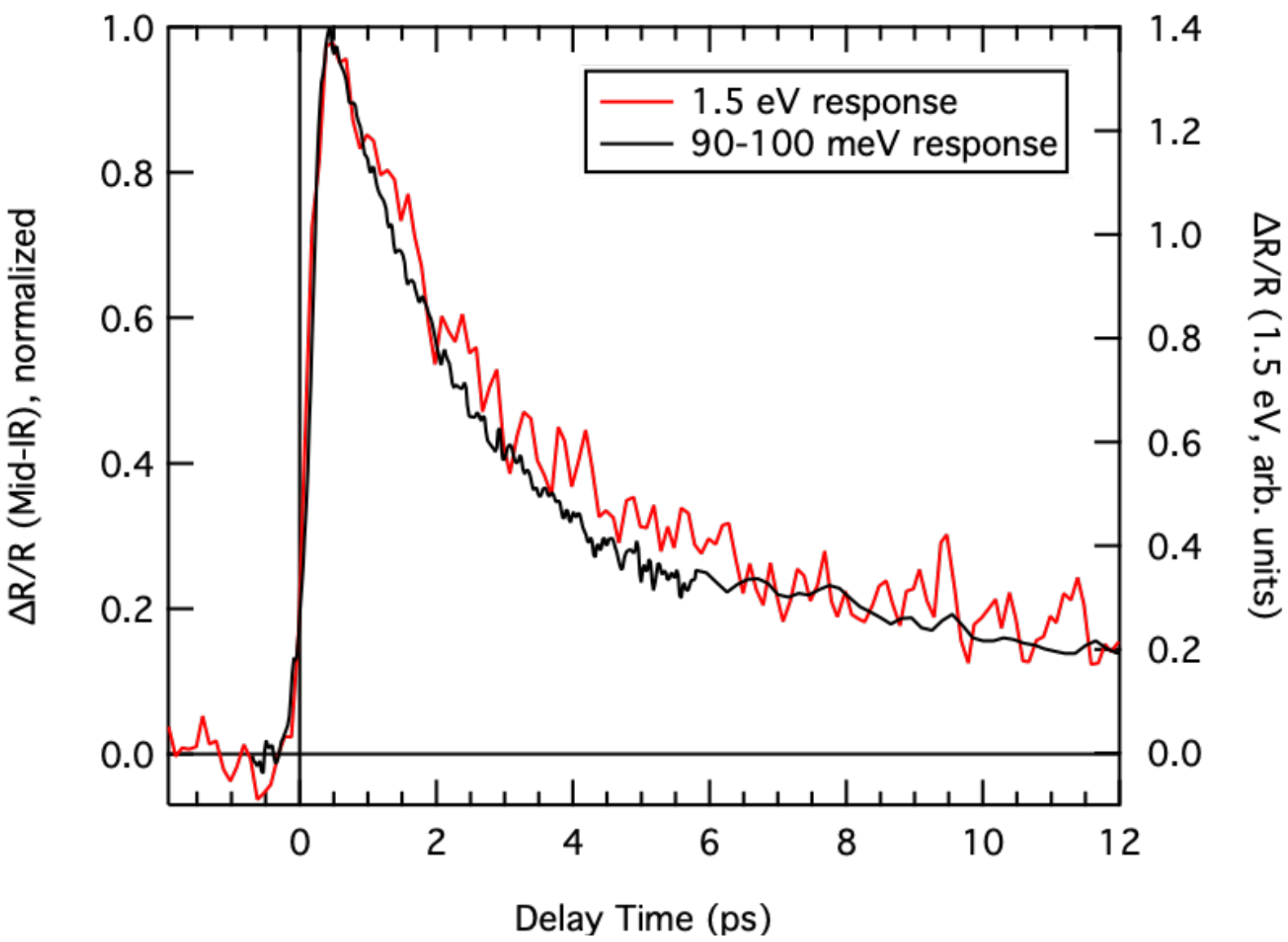

**Fig. S5.** Comparison between the Mid-IR response reported in Ref. ***(37)*** and the response at 1.5 eV at low fluence. The Mid-IR signal shown here is due to the slow (SC) component extracted from the SVD method. This component shows a peak around 90-100 meV associated with the closing of the SC gap ***(37)***.

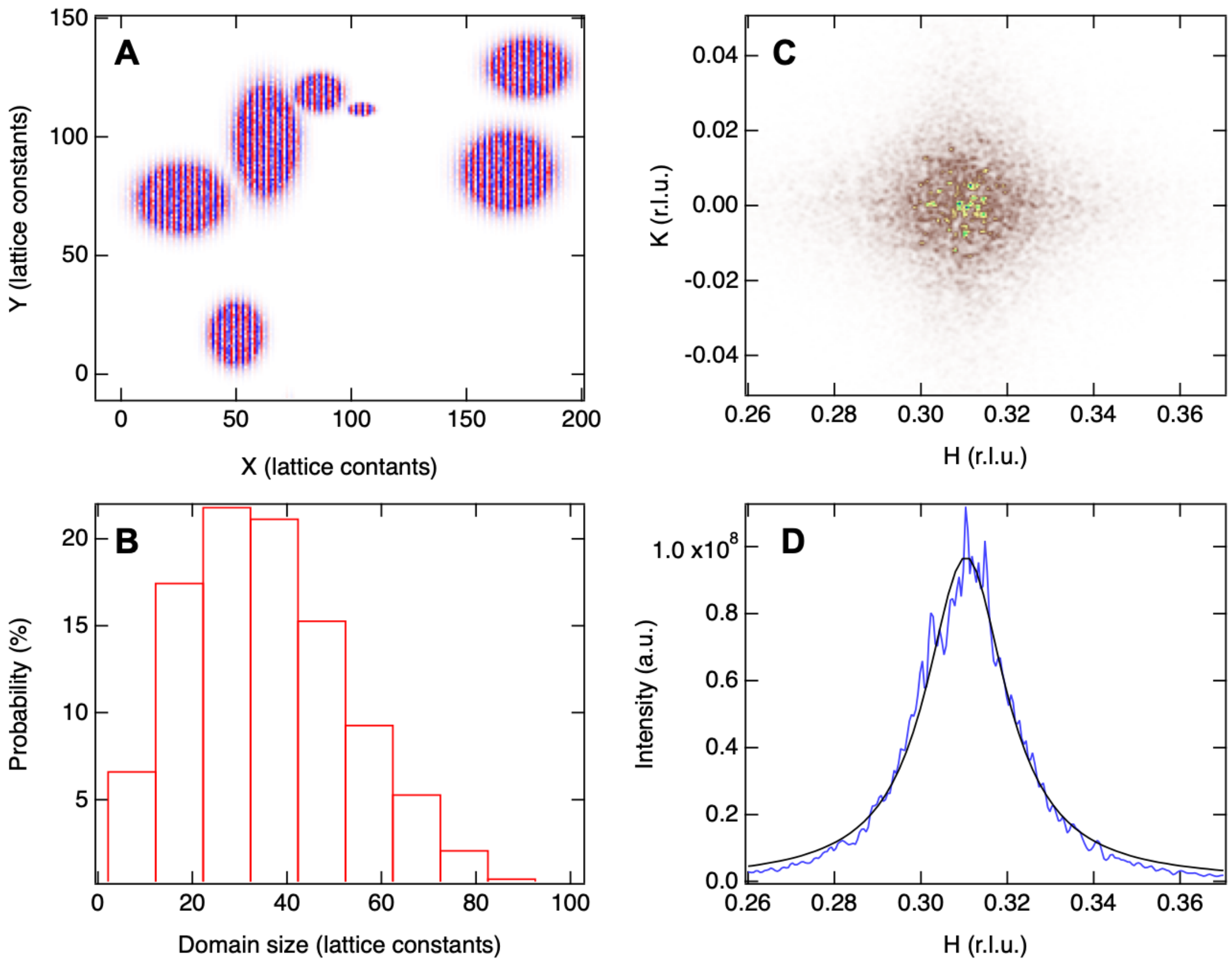

**Fig. S6.** Overview of the simulation of the CDW patterns and x-ray scattering signals. A) Spatial map of the computed charge distribution $\boldsymbol{\rho(x, y)}$, with a typical arrangement of CDW domains. B) Probability distribution of the radial domain size. This relatively broad distribution of domains was considered to optimize the agreement with the experimental data and following micro X-ray diffraction imaging data for the CDW in another cuprate ***(11)***. C) 2D scattering signal map obtained from FFT of A). D) Projection of the scattering map along the H-axis, convoluted with a Gaussian function to include experimental resolution. The curve is fitted with a Lorentzian function and can be directly compared to the experimental data.

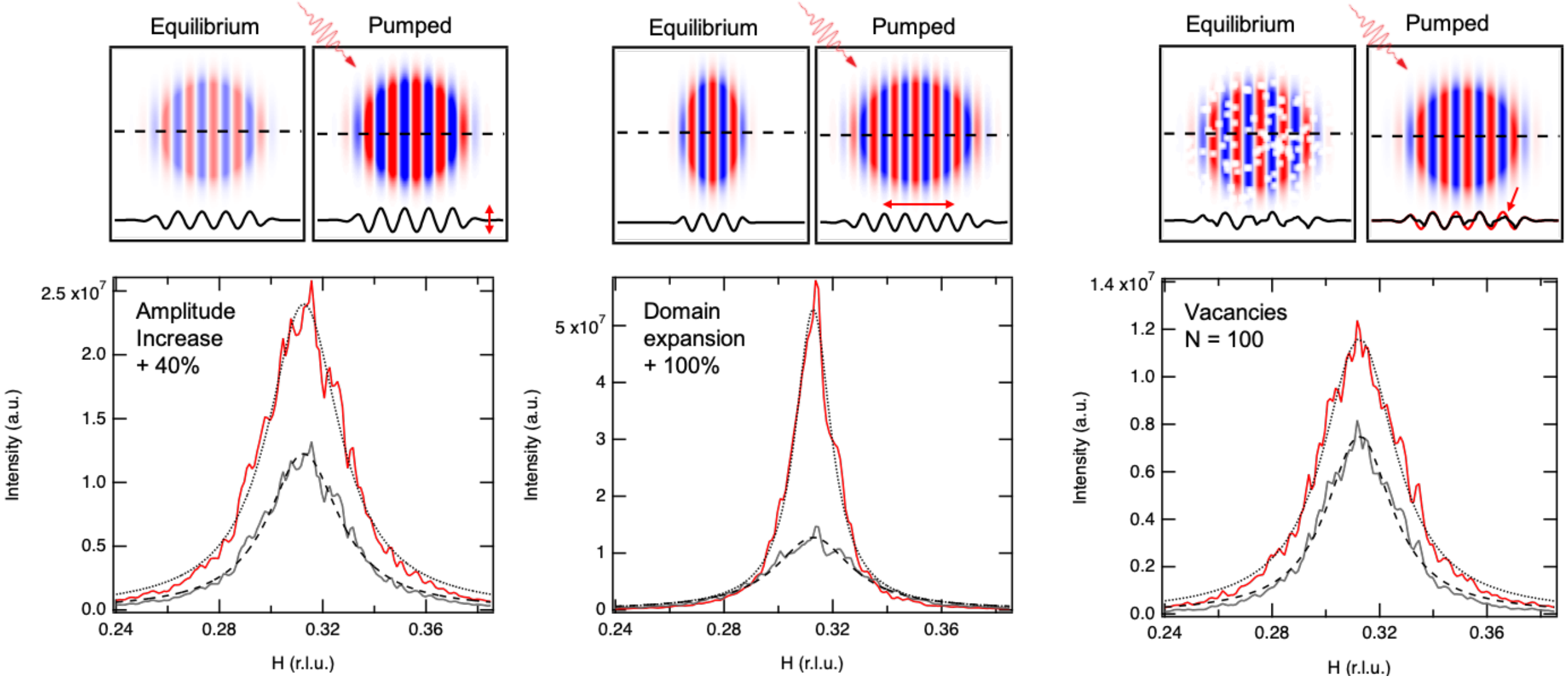

**Fig. S7.** Simulated x-ray scattering signal for the cases reported in Fig. 3 of the main text. A) amplitude increase, as predicted by Ginzburg-Landau theory for coexisting and homogenous orders, B) CDW domain expansion, as expected in the case of phase separation, and C) a distribution of vacancies that is removed by the laser.

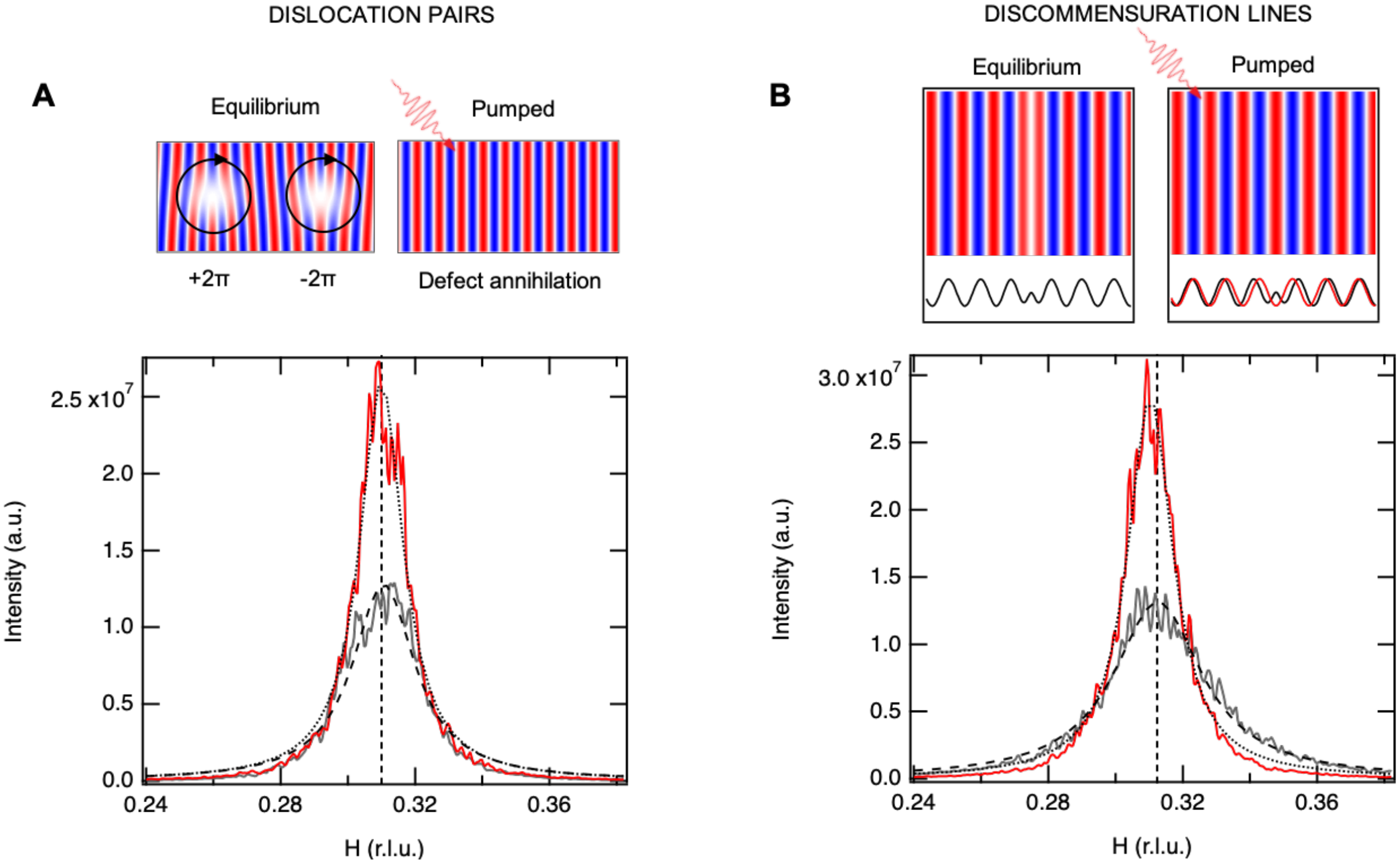

**Fig. S8.** Simulated x-ray scattering signal for the case of topological defects, such as dislocations and discommensuration lines. A) For the case of dislocations, we consider an equal distribution of dislocations characterized by opposite winding phases, +2π and -2π. We introduce these dislocations in the equilibrium state and assume they are completely removed in the pumped state, following laser excitation. Defect annihilation can happen via ultrafast motion of the dislocation toward its counterpart with opposite winding phase, or toward the domain boundary. Upon defect removal in the pumped state, the simulated scattering signal shows an increase of ≈ 40% in both amplitude and correlation length, as well as no significant momentum shift. B) For the case of discommensuration lines, we considered a 1D -π phase jump within the CDW domain. In the pumped state all the defects are removed. The x-ray scattering signal for the pumped state shows a large increase of correlation length (≈ 95%), a ≈ 10% increase of amplitude and a -0.0022 r.l.u. momentum shift.